\documentclass[12pt,preprint]{aastex}

\slugcomment{Accepted for Publication in the Astrophysical Journal}

\shorttitle{Variability in SAO 206462}
\shortauthors{Sitko et al.}

\begin{document}


\title{Variability of Disk Emission in Pre-Main Sequence  and Related Stars. \\
    II. Variability in the Gas and Dust Emission of the Herbig Fe Star SAO 206462}


\author{Michael L. Sitko\altaffilmark{1,2}, Amanda N. Day\altaffilmark{1}, Robin L. Kimes\altaffilmark{1}, Lori C.  Beerman\altaffilmark{1,3}, Cameron Martus\altaffilmark{4}}
\affil{Department of Physics, University of Cincinnati, Cincinnati OH 45221, USA}
\email{sitkoml@ucmail.uc.edu}

\author{David K. Lynch\altaffilmark{1}, Ray W. Russell\altaffilmark{1}}
\affil{The Aerospace Corporation, Los Angeles, CA 90009, USA}
\email{David.K.Lynch@aero.org, Ray.W.Russell@aero.org}

\author{Carol A. Grady}
\affil{Eureka Scientific, Inc., Oakland, CA 94602 \\ and \\ Exoplanets and Stellar Astrophysics Laboratory, \\ Code 667, Goddard Space Flight Center, Greenbelt, MD 20771, USA}
\email{Carol.A.Grady@nasa.gov}

\author{Glenn Schneider}
\affil{Steward Observatory, University of Arizona, Tucson, AZ  85721, USA}
\email{gschneid@email.arizons.edu}

\author{Carey M. Lisse}
\affil{Applied Physics Lab, Johns Hopkins University, 11100 Johns Hopkins Road, Laurel, MD 20723, USA}
\email{carey.lisse@jhuapl.edu}

\author{Joseph A. Nuth}
\affil{Goddard Space Flight Center, Greenbelt, MD 20771, USA}
\email{joseph.a.nuth@nasa.gov}

\author{Michel Cur\'{e}}
\affil{Departamento de F\'{\i}sica y Astronom\'{\i}a, Universidad de 
Valpara\'{\i}so, Avda. Gran Breta\~na 1111, Valpara\'{\i}so, Chile}
\email{michel.cure@uv.cl}

\author{Arne A. Henden}
\affil{American Association of Variable Star Observers, 49 bay State Road, Cambridge, MA 02138, USA}
\email{arne@aavso.org}

\author{Stefan Kraus}
\affil{Dept. of Astronomy, University of Michigan, Ann Arbor, MI 48109, USA}
\email{stefankr@umich.edu}

\author{Veronica Motta}
\affil{Departamento de F\'{\i}sica y Astronom\'{\i}a, Universidad de 
Valpara\'{\i}so, Avda. Gran Breta\~na 1111, Valpara\'{\i}so, Chile}
\email{vmotta@dfa.uv.cl}

\author{Motohide Tamura}
\affil{National Astronomical Observatory of Japan, 2-21-1 Osawa, Mitaka, Tokyo 181-8588, Japan}
\email{motohide.tamura@nao.ac.jp}

\author{Jeremy Hornbeck}
\affil{Dept. of Physics \& Astronomy, University of Louisville, Louisville, KY 40292, USA}
\email{jbhorn02@louisville.edu}

\author{Gerard M. Williger}
\affil{Dept. of Physics \& Astronomy, University of Louisville, Louisville, KY 40292, USA}
\email{williger@pha.jhu.edu}

\and

\author{Dino Fugazza}
\affil{INAF, Osservatorio Astronomico di Brera, via E. Bianchi 46, 23807 Merate, Italy}
\email{dino.fugazza@brera.inaf.it}


\altaffiltext{1}{Guest Observer, Infrared Telescope Facility}
\altaffiltext{2}{Also Space Science Institute, Boulder}
\altaffiltext{3}{Now at the University of Washington}
\altaffiltext{3}{Now at the University of Alaska}




\begin{abstract}

We present thirteen epochs of near-infrared (0.8-5 \micron{}) spectroscopic observations of the pre-transitional, ``gapped'' disk system in SAO 206462 (=HD 135344B). In all, six gas emission lines (Br$\alpha$, Br$\gamma$, Pa$\beta$, Pa$\gamma$, Pa$\delta$, Pa$\epsilon$, and the 0.8446 $\mu$m line of O I) along with continuum measurements made near the standard J, H, K, and L photometric bands were measured. A mass accretion rate of approximately 2 x 10$^{-8}$ M$_{\sun}$ yr$^{-1}$ was derived from the Br$\gamma$ and Pa$\beta$ lines. However, the fluxes of these lines varied by a factor of over two during the course of a few months. The continuum also varied, but by only $\sim$30\%, and even decreased at a time when the gas emission was increasing. The H I line at 1.083 $\mu$m was also found to vary in a manner inconsistent with that of either the hydrogen lines or the dust. Both the gas and dust variabilities indicate significant changes in the region of the inner gas and the inner dust belt that may be common to many young disk systems. If planets are responsible for defining the inner edge of the gap, they could interact with the material on time scales commensurate with what is observed for the variations in the dust, while other disk instabilities (thermal, magnetorotational) would operate there on longer time scales than we observe for the inner dust belt. For SAO 206462, the orbital period would likely be 1-3 years. If the changes are being induced in the disk material closer to the star than the gap, a variety of mechanisms (disk instabilities, interactions via planets) might be responsible for the changes seen. The He I feature is most likely due to a wind whose orientation changes with respect to the observer on time scales of a day or less. To further constrain the origin of the gas and dust emission will require multiple spectroscopic and interferometric observations on both shorter and longer time scales that have been sampled so far.

\end{abstract}




\section{Introduction}

Planetary systems are born from disks of gas and dust surrounding young stars. The rate of discovery of planetary systems demonstrates that this process is efficient. These disks generally become cleared in $<$10 Myr, with the exception of gas-poor debris disks regenerated by collisional cascades. For decades it was realized that planet formation and the photoerosion of disk material would clear this disk material on this same time scale, and in 1989 the first evidence of disks with cleared inner zones was presented by \citet{strom89}, who referred to these as \textit{disks in transition}, from which the current term ``transitional disks'' is derived. Since that time more detailed spectral energy distributions (SEDs) obtained with ground-based telescopes and the \textit{Spitzer Space Telescope} and near-infrared and millimeter interferometry have provided astronomers with the tools needed to investigate the structure and dynamics of these systems more completely. The prototypical transitional disk, that of TW Hya, contains a nearly cleared inner zone that has been determined first from its SED \citep{calvet02} and confirmed interferometrically at 2~\micron{} \citep{eisner06}, 10~\micron{} \citep{ratzka07}, and 7 mm \citep{hughes07}.

It has recently become apparent that some of these transitional disk systems contain a separate inner belt of dust, making them ``gapped'' disks. Currently one of the best-studied of these objects is SAO 206462 = HD 135344B\footnote{Originally, HD 135344 referred to  both stars in the binary system, an A star with no appreciable emission from a circumstellar disk, and an F star, which has been the star usually referred to alone as ``HD 135344'' in much of the literature. However this has often led to confusion, with some data on one star being attributed to the other. The F star is now generally referred to as HD 135344B to make it clear it is not the A star, but we prefer to minimize any confusion further by using its SAO designation (the A star is SAO 206463).}, an F4Ve star \citep{dunkin97} whose distance is uncertain, but is likely between 140 pc  and 170 pc \citep{grady09}. For a distance of 160 pc, the likely stellar mass and radius are $\sim$1.6 M$_{\sun}$ and $\sim$2.3 R$_{\sun}$, respectively (see Appendix A). Here, both the spectral energy distribution (SED) \citep{grady09} and interferometry \citep{fedele08} indicate the presence of a double-disk structure. Such a gap cannot be due to photodestruction, which would clear the disk``inside-to-outside''. But a gap in the disk is the expected signature of clearing by a planetary-mass object orbiting the star in the cleared zone. This phase of disk evolution can lead to planet-cleared inner zones such as that seen in TW Hya and other transitional disks is now referred to as a ``pre-transitional'' disk phase  \citep{espaillat07}.

Models of the SED by \citet{grady09} indicated that the inner belt of dust material must be radiating at temperatures in excess of 1400 K, suggesting that it was composed primarily of ``super-refractory'' material, as silicates would not be expected to survive at those temperatures, and no significant emission feature at 10~\micron{} was visible in its spectrum. Similar materials have been implicated for the hottest dust in the Herbig Ae star HD 163296 (MWC 275) by \citet{benisty10}. In the case of SAO 206462, \citet{grady09} suggested that the hot dust belt was dominated by carbonaceous dust, as had been previously suggested for the disk of $\beta$ Pictoris \citep{roberge06}, the debris belt of HR 4796A \citep{debes08}, and which may even dominate the present zodiacal dust content of the inner solar system \citep{nesvorny10}. 

Another unusual characteristic of the inner dust belt of SAO 206462 found by \citet{grady09} was its apparent photometric \textit{variability}. Inner disk variability has been reported for HD 163296 and HD 31648 (MWC 480) \citep{sitko08}, the 10 $\mu$m silicate band of DG Tau  \citep{woodward04,sitko08,bary09} and XZ Tau \citep{bary09}, as well as the inner and outer disk regions of a number of transitional and pre-transitional disks observed with the \textit{Spitzer Space Telescope}  \citep{muzerolle09,espaillat11}.

SAO 206462 also exhibited changes in the line strengths of the hydrogen Paschen and Brackett lines \citep{grady09}. These lines are usually attributed to the accretion column onto the star, from which mass accretion rates are derived \citep{muzerolle98a,muzerolle98b,muzerolle01}, so changes in the line strengths would indicate variability in the gas accretion rate. For Br$\gamma$ the line strength dropped by 50\% over a span of two months in 2007. With only three observations of the line strength, it was impossible to decide if the changes were:

$\bullet$ simply stochastic fluctuations in the long-term decline of gas accretion expected in aging disk systems (``sputtering out'').

$\bullet$ fueled by some sort of driving mechanism, such as disk instabilities (thermal, magnetorotational). 

$\bullet$ driven by a planet that might be responsible for clearing the gap in the disk.

To begin to address this aspect of pre-transitional disk evolution, we observed SAO 206462 using the SpeX spectrograph on NASA's Infrared Telescope Facility (IRTF) with a monthly cadence between February and July 2009, followed by an additional 4 observations in 2011. Combined with the earlier three epochs, these provided 13 epochs of spectroscopic data on SAO 206462, from which 7 emission lines and 4 continuum bands (roughly coincident with the standard J, H, K, and L' bandpasses) were measured. In 2011, VRIJHK photometry was also performed using the Rapid Eye Mount facility. Mass accretion rates were determined for all 13 epochs. In this paper we discuss the character of the line and continuum variability, the gas accretion rates derived from the Pa$\beta$ and Br$\gamma$ lines of hydrogen, the behavior of the He I line at 1.083 $\mu$m, and place some constraints on the possible mechanisms that may be responsible for the observed variations.

\section{Observations \& Spectral Processing}

\subsection{SpeX Cross-Dispersed Observations}

The SpeX observations were made using the cross-dispersed (hereafter XD) echelle gratings in both short-wavelength mode (SXD) covering 0.8-2.4 \micron{} and long-wavelength mode (LXD) covering 2.3-5.4 \micron{} \citep{rayner03}. Data in both XD modes were obtained for all but one epoch. These observations were obtained using a 0.8 arcsec wide slit, corrected for telluric extinction, and flux calibrated against a variety of A0V calibration stars observed at airmasses close to that of SAO 206462. The SpeX observations are summarized in Table 1. The data were reduced using the Spextool software \citep{vacca03,cushing04} running under IDL. 

Due to the light loss introduced by the 0.8 arcsec slit used to obtain the SXD and LXD spectra, changes in telescope tracking and seeing between the observations of SAO 206462 and a calibration star may result in merged SXD or LXD spectra with a net zero-point shift compared to their true absolute flux values. We used a variety of techniques to check for any systematic zero-point shift in the absolute flux scale, as discussed below in greater detail. These included using the low dispersion prism in SpeX with a 3.0 arcsec wide slit, JHK photometry with the SpeX guider, spectrophotometry with The Aerospace Corporation's  Broad-band Array Spectrograph System (BASS) with a 3.4 arcsec aperture and 3-13 $\mu$m spectral coverage, BVR${_c}$I$_{c}$ photometry obtained independently using the CCD camera of the \textit{Sonoita Research Observatory} in Arizona, part of the telescope network of the American Association of Variable Star Observers (AAVSO)\footnote{www.aavso.org}, and VRIJHK photometry obtained with the Rapid Eye Mount (REM), La Silla, Chile\footnote{ www.rem.inaf.it}.

\subsection{SpeX Prism Observations}

On many nights we observed SAO 206462 and its A0V calibration star using the low-dispersion Prism and a slit 3.0 arcsec wide. To avoid saturation of the detector and minimize the wavelength calibration arc line blends, the flat-field and wavelength calibration exposures required a narrower slit, and the 0.8 arsec slit was used.

On most nights with seeing better than 1.0 arcsec, very little light hit the slit jaws, and individual exposures showed minimal scatter. However, star motion on the detector can introduce small wavelength shifts, which will result in under-corrected or over-corrected telluric bands with a characteristic ``P Cygni'' shape. In the prism spectra shown here, these were not in evidence, however. 

\subsection{SpeX Near-Infrared Photometry}

On a few nights, images were obtained with the J, H, and K filters of the SpeX guide camera (``Guidedog''). The Mauna Kea filter set is described by \citet{simons02}, \citet{tokunaga02}, and \citet{tokunaga05}, and the data obtained were calibrated using the JHK photometric standard stars of \citet{leggett06} for that filter set. Aperture photometry on each image was performed using the ATV.pro routine running under IDL. 

\subsection{BASS Spectrophotometry}

We observed SAO 206462 with BASS on 9 July 2007 and 16 July 2009 (UT). BASS uses a cold beamsplitter to separate the light into two separate wavelength regimes. The short-wavelength beam includes light from 2.9-6 $\mu$m, while the long-wavelength beam covers 6-13.5 $\mu$m. Each beam is dispersed onto a 58-element Blocked Impurity Band (BIB) linear array, thus allowing for simultaneous coverage of the spectrum from 2.9-13.5 $\mu$m. The spectral resolution $R = \lambda$/$\Delta\lambda$ is wavelength-dependent, ranging from about 30 to 125 over each of the two wavelength regions \citep{hackwell90}. The BASS observations are summarized in Table 2.

\subsection{Visible Wavelength Photometry - AAVSO}

SAO 206462 was observed twice per night on approximately 75 nights in 2009 over the date interval JD 2454876 through JD 2455028 using the robotic 35cm telescope at \textit{Sonoita Research Observatory}. Observations were made in the Johnson/Cousins BVR$_{c}$I$_{c}$ passbands, with the measurements tied to that system using a complement of observed fields containing photometric standard stars. The photometry used digital apertures of approximately 13 arcsec diameter, along with an ensemble of comparison stars (typically 5 stars per measure). 

\begin{figure}
\center
\plottwo{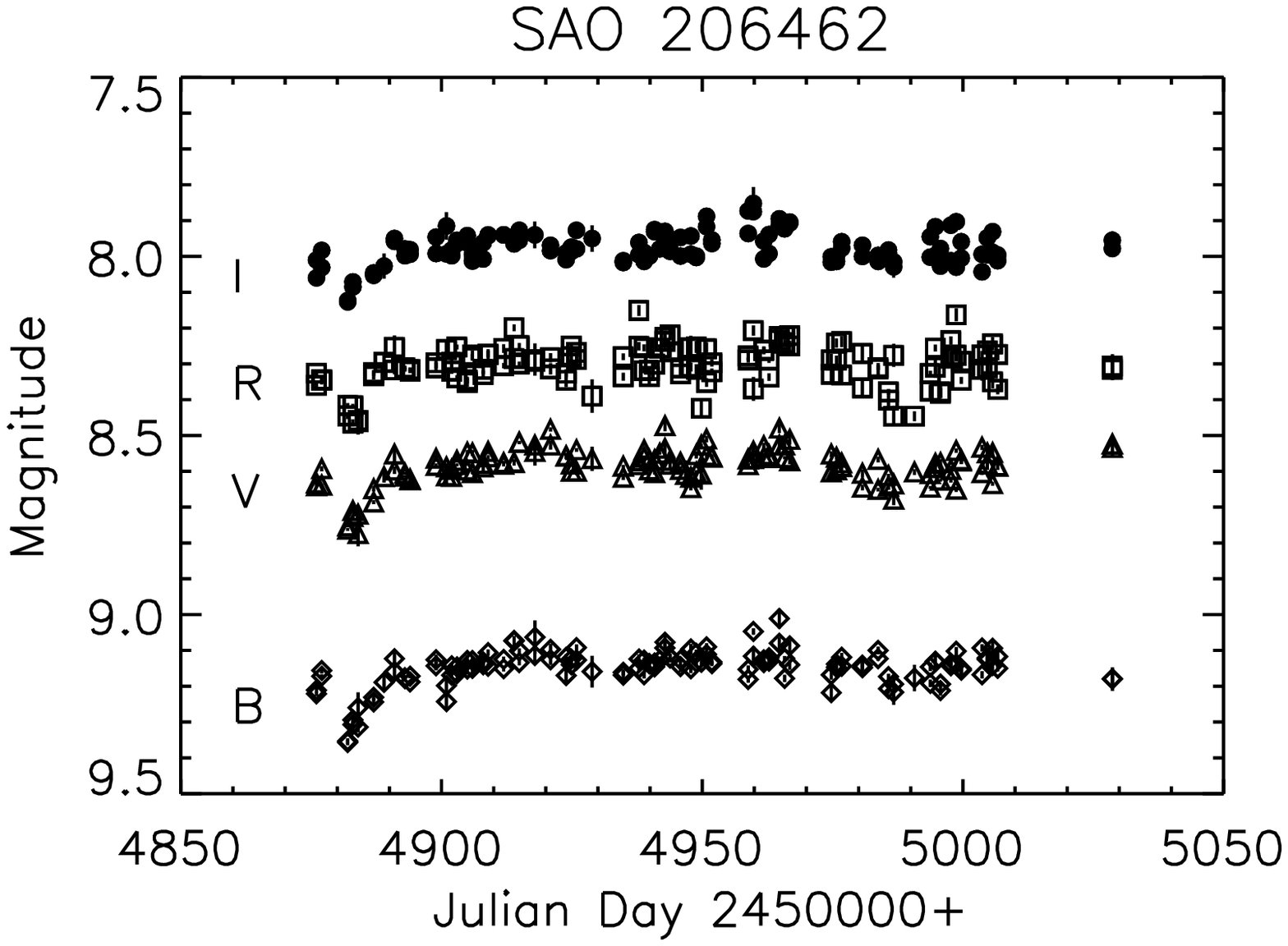}{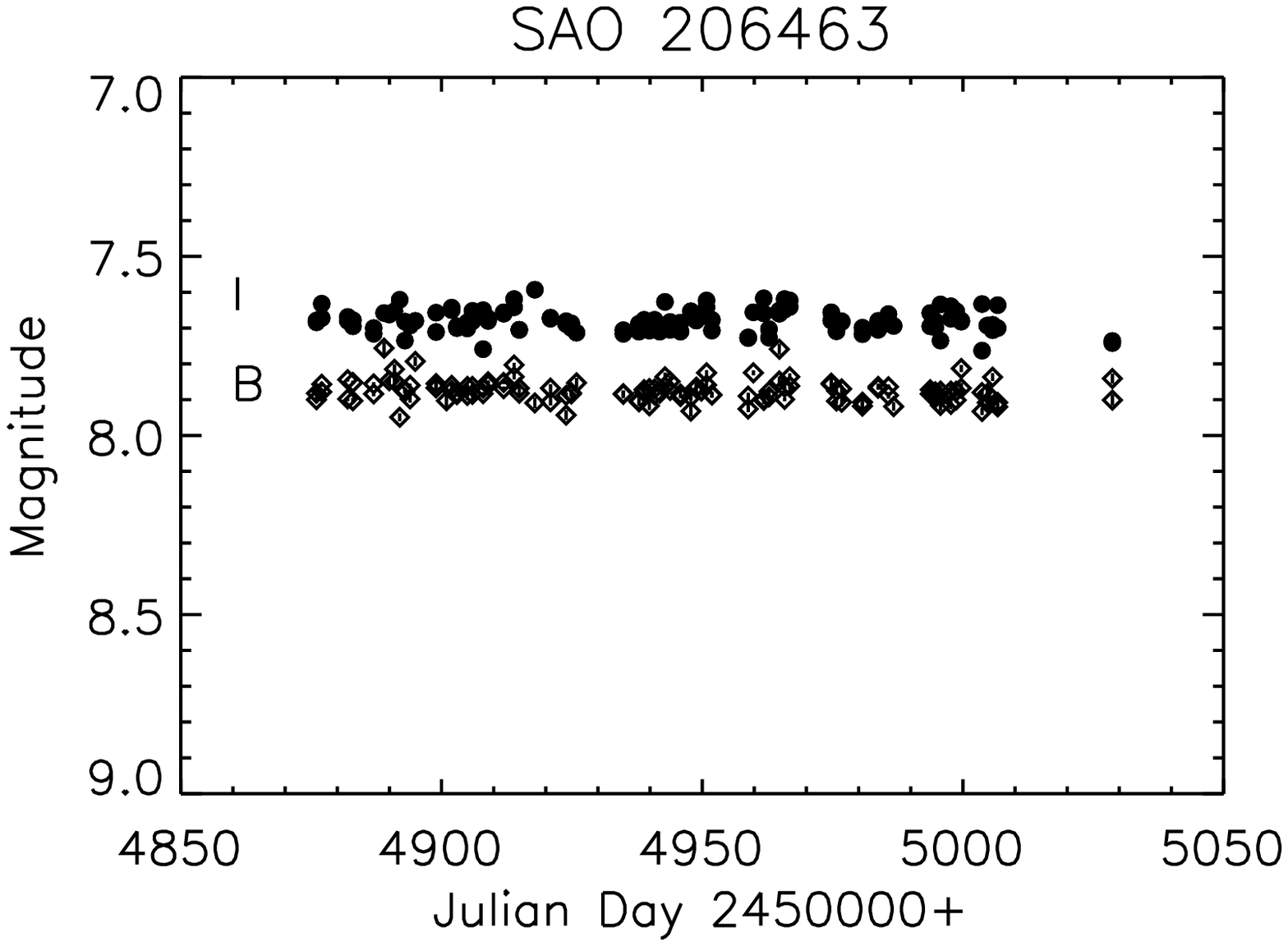}
\caption{Left: BVRI photometry of SAO 206462 (HD 135344B), obtained in 2009 with the 35-cm telescope of the \textit{Sonoita Research Observatory}, operated in part by the AAVSO. The point-to-point scatter is dominated by scintillation and atmospheric changes. Longer-term modulation is possibly due to star spot activity. The drop in brightness seen near JD=2454880 days is consistent with the lower flux level derived from the SpeX Prism data in 2009 Feb. 18. Right: Photometry of SAO 206463 (HD 135344A), the A0V star located 22 arcsec from SAO 206462, and based on the same photometric calibration. For clarity, only the data in the B and I filters are shown. The flux drop seen in SAO 206462 is not seen here, conforming its reality for the other star. \label{fig:magslightcurve}}
\end{figure}

At -37 degrees declination, SAO 206462 is a difficult object from Arizona, transiting at 2.8 airmasses.  While careful attention was paid to first and second order extinction, and all results are transformed, the majority of the scatter for this bright star results from atmospheric changes and scintillation. The BVR$_{c}$I$_{c}$ light curves are shown in  Fig.~\ref{fig:magslightcurve}. For the final light curves, nights where the two measurements disagreed by more than a few hundredths of a magnitude, or which resulted from the use of only one calibration star, were rejected.

\subsection{Visible \& Near-Infrared Photometry - REM}

SAO 206462 was also observed in 2011 March and May with the REM. This robotic facility simultaneously observed the same field in VRIJHK. Because many of these data were obtained on nights where the sky transparency was variable, SAO 206462 was measured differentially with respect to SAO 206463, only 22 arcsec away, which seems photometrically stable. Standard JHK magnitudes for SAO 206463 were taken from 2MASS, while the VRI magnitudes were taken from the mean values of the AAVSO observations obtained in 2009.

\subsection{Absolute Flux Calibration of the SpeX Data}

\begin{figure}
\center
\plottwo{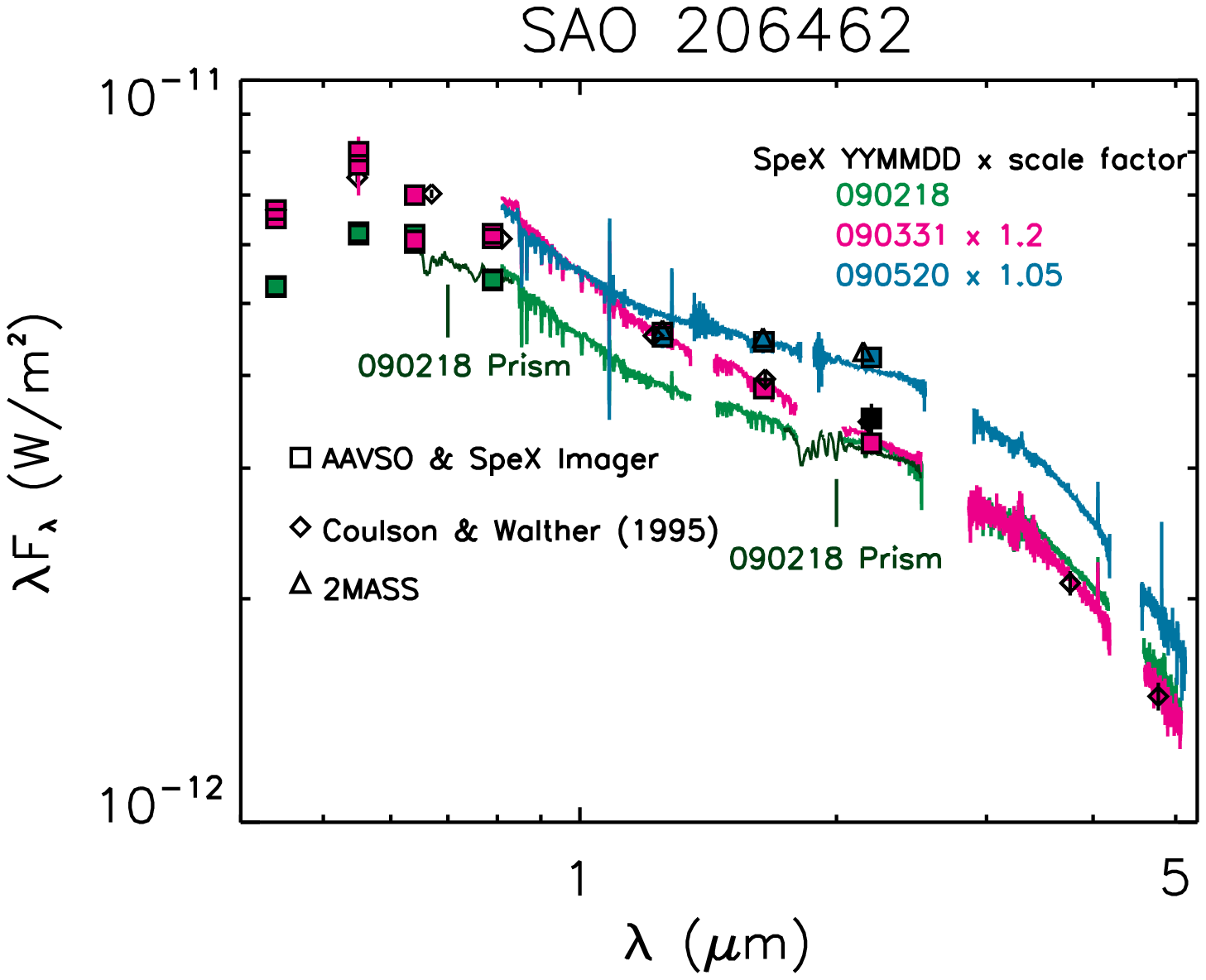}{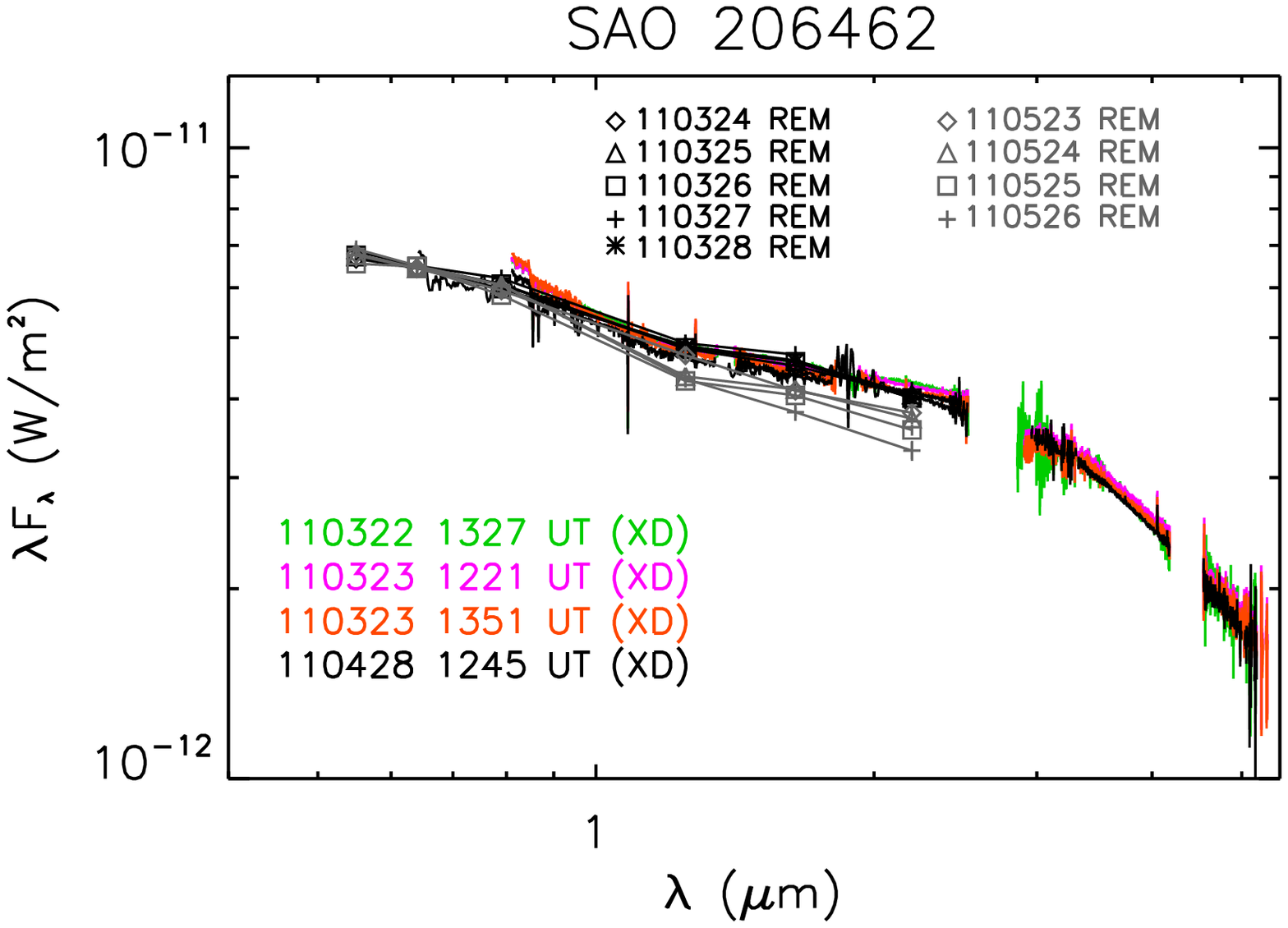}
\caption{Left: SAO 206462 observed on 3 epochs in 2009, illustrating the absolute calibration process.  Also shown are archival 2MASS data, which seem to have been obtained during a state similar to that to 20 May, and BVRIJHKLM photometry from \citet{coulson95}, when SAO 206462 was in a ``steep spectrum'' state similar to that of 31 March. Right: The SpeX XD data obtained in 2011 March were normalized using REM photometry obtained within 1-2 days of the spectra. In the five days of photometry obtained in March, SAO 206462 was photometrically stable. Additional photometry obtained in May indicated that the emission from the inner dust disk had declined (the beginning of another ``steep spectrum'' event), but the VRI emission levels indicated that this was not due to changes in the stellar photospheric brightness at these wavelengths.} \label{fig:3epochs}
\end{figure}

In Fig.~\ref{fig:3epochs} we illustrate the use of the normalization process on three epochs of data obtained in 2009 and three epochs in 2011. For the SpeX XD data obtained on 18 February 2009 we also show the Prism data obtained with the 3.0 arcsec slit. In this case, the XD data is almost identical in flux level as the Prism data, and no scaling was applied to get the absolute flux levels. (The Prism data between 0.8 - 1.6 $\mu$m were saturated and are not plotted.) Also shown in the same color as the XD data is the AAVSO photometry from 19 February. These agree with both the Prism and unscaled XD data. For 31 March, JHK imaging with the SpeX guider and AAVSO photometry from 30 March were used to determine the scaling. Both the imaging photometry and the XD spectrum indicate a steeper SED on this date than on 18 Feb., being brighter at the shorter wavelengths but comparable at longer wavelengths. data. For 20 May, JHK imaging suggest a shallower spectrum than 31 March, being brighter at the longer wavelengths.

For the 2011 March 22 \& 23 (UT) spectra, we were fortunate to obtain VRIJHK photometry with REM (24-28 March UT) in the days immediately following the SpeX observations, and during a time when SAO 206462 was photometrically stable. These data were used to set the scaling of these three spectra.

\section{Line and Continuum Strengths}

\subsection{Continuum Emission}

In order to extract the net energy flux in each line, a model was constructed to reproduce the background continuum emission of the stellar photosphere and the dust emission of the inner dust belt. For the stellar photosphere we use the SpeX data on HD 87822, an F4V star in the IRTF/SpeX spectral library \citep{cushing04,rayner09}. Because the slit widths were different (0.3 arcsec for HD 87822 versus 0.8 arcsec for SAO 2086462) and differences in seeing (affecting mostly SAO 206462) we smoothed the spectrum of HD 87822 with a gaussian function to match as nearly as possible the photospheric features in each spectrum of SAO 206462 between 0.8-1.0 $\mu$m, where the photosphere dominates the total flux.

\begin{figure}
\center
\plotone{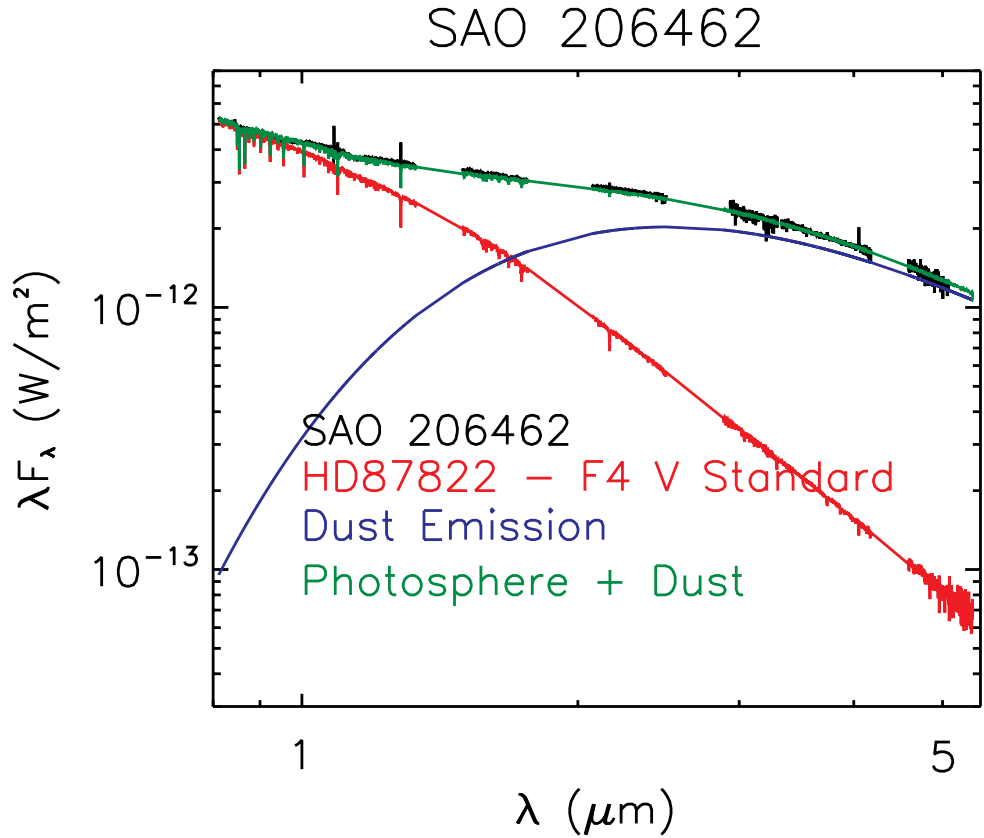}
\caption{Model for SAO 206462. The SED consists of the underlying photospheric emission of the star plus thermal emission by the inner dust ring.The relative strengths of the two components were adjusted to match the observed continuum emission of the star plus disk system. \label{fig:model}}
\end{figure}

Once a suitable match was made a modified blackbody representing the emission of the inner dust belt was added by multiplying a planck function by a wavelength-dependent emissivity:

F$_{BB}$=B($\lambda$,T) $\cdot$ $\lambda^{\beta}$

\noindent where T is the temperature and $\lambda$ is the wavelength. The actual shape of the dust emission will in general be a complex function of the geometry, radial and vertical dust density, and grain optical properties of the emitting region. The model here is not intended to be a physical one, but an approximation to the effects of having a range of grain temperatures and wavelength-dependent grain emissivities contributing to the underlying continuum for the purpose of removing everything other than the line fluxes\footnote{This implies that any underlying gaseous continuum is also being removed, to first order.}. The model continuum flux was then subtracted. As an example, the model for the data obtained on 10 July 2009 are shown in Fig.~\ref{fig:model}.

\begin{figure}
\center
\includegraphics[scale=.80]{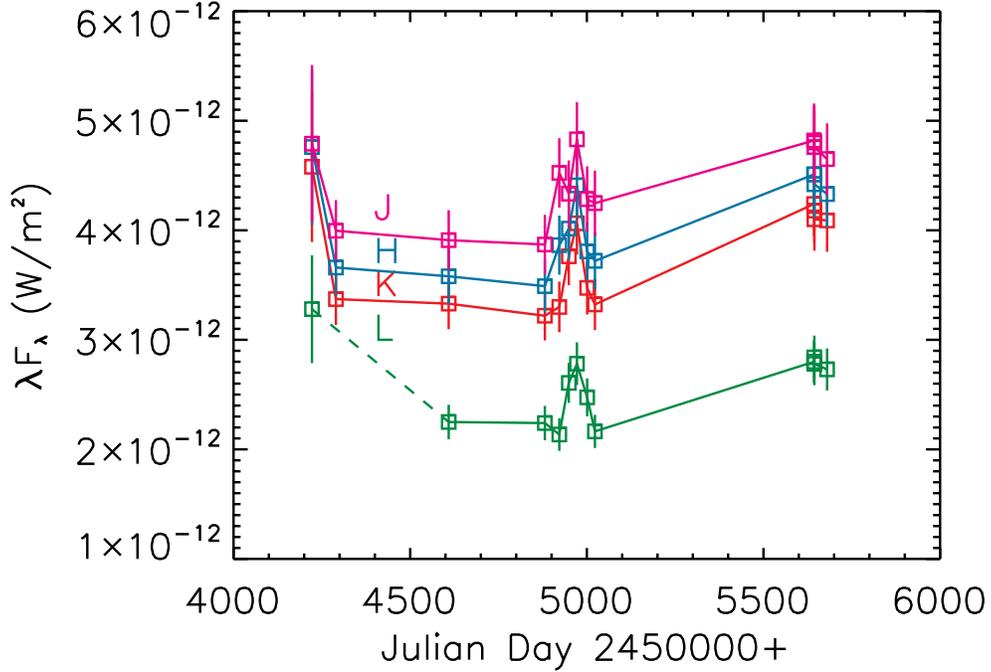}
\caption{The time variability  of the near-IR continuum for all 13 epochs. The outburst in 2009 seems to have appeared first in the J-band, and is similar to the ``high-J'' state described by \citet{grady09}. We have set the uncertainties on the data at all but the first epoch at 7\%, which are propagated into the line flux measurements; the agreement among the various flux-normalization sets is often much better. The AAVSO photometry and SpeX Prism data on 18 February agree to better than this precision. The uncertainty in the first epoch was set to a generous 15\% because no independent flux calibration data was available for it. As no LXD spectrum was obtained on 8 July 2007, there is no L-band flux plotted for that date.\label{fig:SpeX_lightcurve_JHK}}
\end{figure}

Monochromatic fluxes roughly coincident with the JHKL bandpasses were also extracted from the flux-calibrated SpeX data by integrating over the spectra between 1.20-1.30~\micron{}, 1.60-1.70~\micron{}, 2.109-2.30~\micron{}, and 3.42-4.12~\micron{}. The light curves derived in the J, H, K, and L bands is shown in Fig.~\ref{fig:SpeX_lightcurve_JHK}.

Because the models were not able to produce a pseudocontinuum that matched the data in every spectral order at every line to be extracted, the model was adjusted locally using a vertical scaling until the $\chi^{2}$ of the difference in the continuum nearby (but outside) the line was minimized. For the majority of the lines on all nights, these corrections to the scaling were less than 2\% of the initial model continuum level. 

\subsection{Line Emission}

\begin{figure}
\center
\plotone{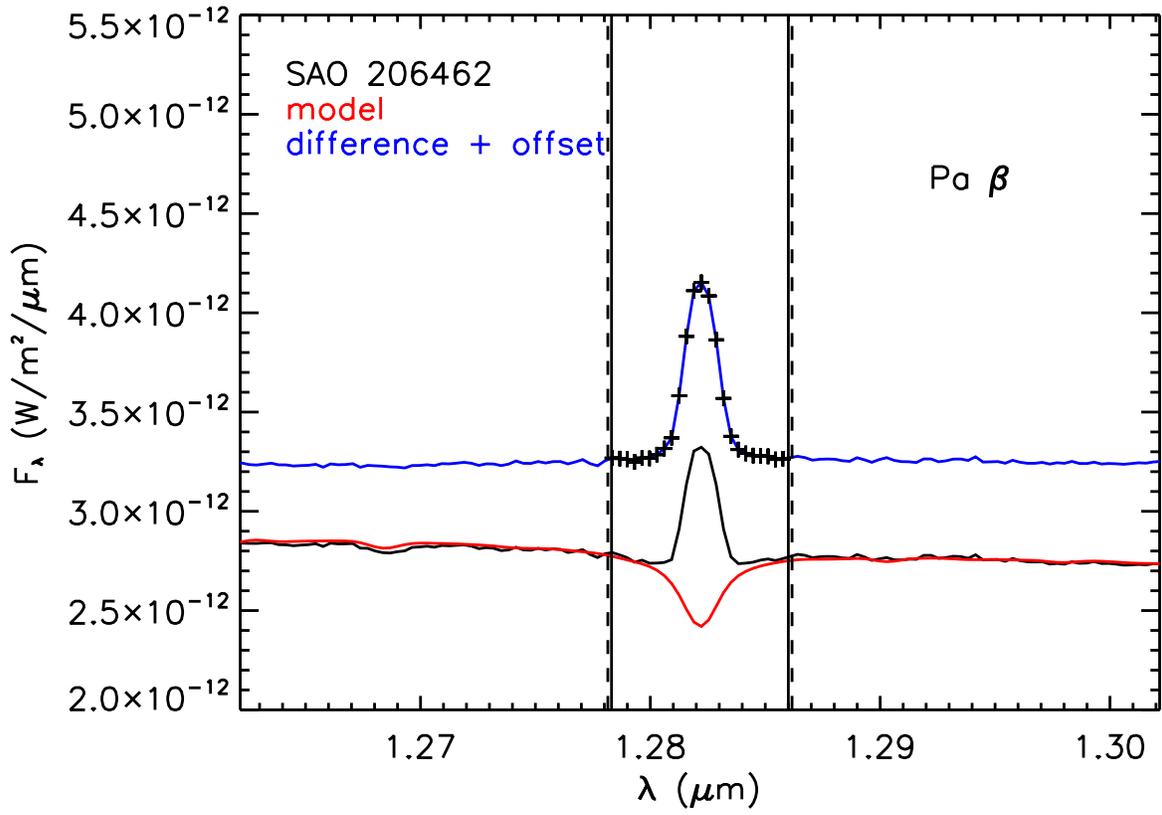}
\caption{Line extraction for Paschen $\beta$ for 10 July 2009 UT. Here the difference between the observations and the model is offset for clarity. The vertical lines and the plus symbols indicate the wavelength range over which the net line strength was measured. \label{fig:Pa_beta}}
\end{figure}

\begin{figure}
\center
\plotone{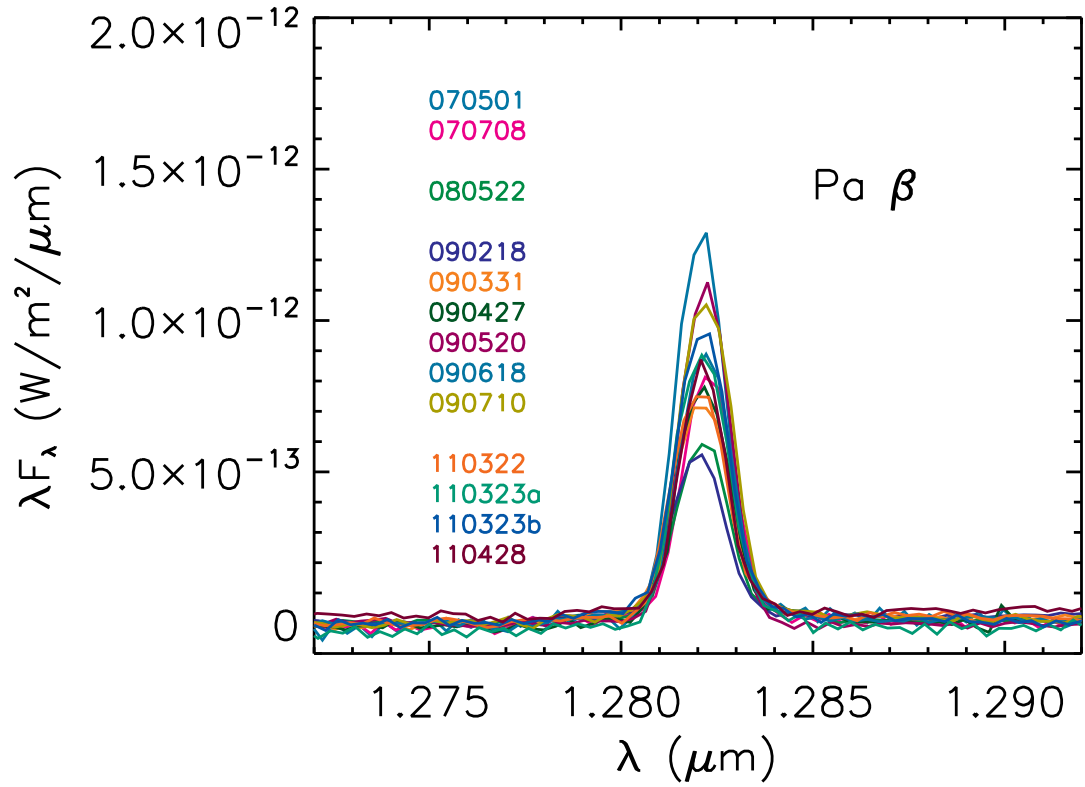}
\caption{Flux-calibrated continuum-subtracted Paschen $\beta$ for all 13 epochs .The ``jitter'' in the central wavelength is likely dominated by a combination of the spectral sampling and uncertainties in the wavelength calibration.  \label{fig:line-PaB}}
\end{figure}

The line strengths were then extracted by subtraction of the adjusted model from the flux-calibrated data, and the net line flux calculated by integrating over the flux-calibrated continuum-subtracted line profile. In Fig.~\ref{fig:Pa_beta} we show an example of the extraction process for the Pa$\beta$ line on 10 July 2009. In Fig.~\ref{fig:line-PaB}, the resulting continuum-subtracted Pa$\beta$ for all 13 epochs is shown. 

The line fluxes extracted in this manner for Pa$\epsilon$, Pa$\delta$, Pa$\gamma$, Pa$\beta$, Br$\gamma$  and Br$\alpha$ and the O I line at 0.8446 $\mu$m  for each epoch are shown in Table 3. While line strengths were also extracted for the Ca II triplet near 0.85 $\mu$m, and the Pa 14, Pa 12, Pa 11, Pa 10, and Pa 9 lines, these were very weak, in many cases only partially filling in the absorption featured of the photosphere. Because SAO 206462 is a rapidly rotating star, its photosphere cannot be approximated accurately by using a spectrum characterized by a single effective temperature and surface gravity, so only the strongest emission lines that are located where the photospheric lines are weak, such as Pa$\epsilon$, Pa$\delta$, Pa$\gamma$, Br$\gamma$  and Br$\alpha$ are useful in the analysis of this star. O I is a different case, as it sits on top of weak circumstellar emission and photospheric absorption of Pa 18. However, we retained this line, as it is being produced at least in part (and perhaps primarily)  by H I Ly$\beta$ Bowen fluorescence, and it might provide some measure of the strength of Lyman line emission, which is inaccessible from the ground.  

\begin{figure}
\center
\plottwo{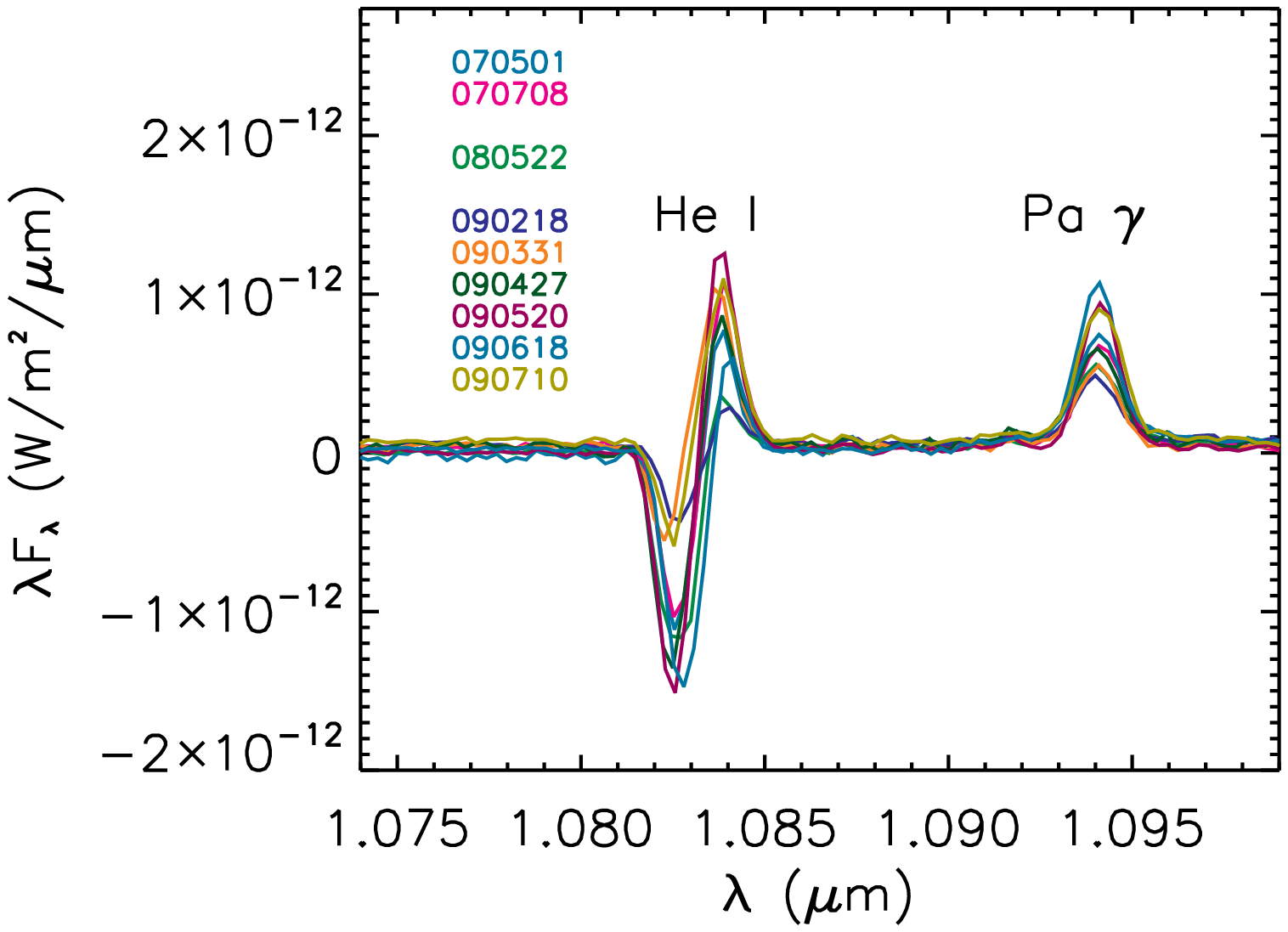}{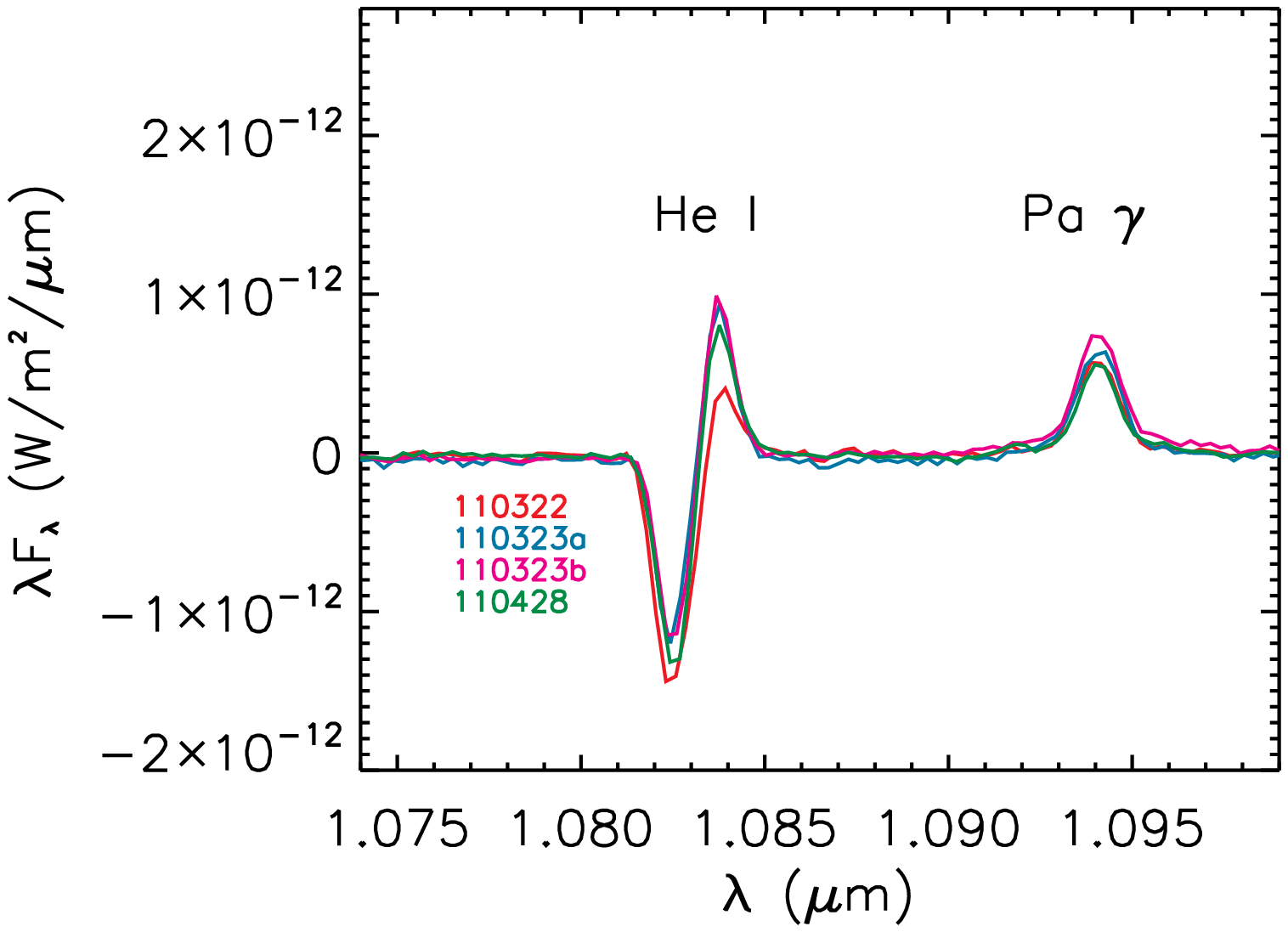}
\caption{Left: Flux-calibrated continuum-subtracted He I 1.08 $\mu$m and Paschen $\gamma$ lines for all 9 epochs between 2007 and 2009 .  As in Fig. 6, the Pa $\gamma$ line exhibits a small amount of epoch-to-epoch shifts in central wavelength, but here is is less than one resolution element. In contrast, the He I line, with a wind-dominated profile, exhibits much greater variability. Right: the same lines during 2011. The two sets of data from 23 March were obtained only 2 hours apart, and are nearly identical. The He I line changed significantly between 22 and 23 March.  \label{fig:Pa_gamma}}
\end{figure}

We also measured the He I line at 1.083 $\mu$m, which exhibits a combination of blended emission and absorption, with a ``P Cygni'' type of line profile, indicative of mass outflow, usually attributed to some form of wind. For this line, the strengths of the longer-wavelength emission and shorter-wavelength absorption were measured separately. The continuum-subtracted He I lines for all epochs, along with those of the adjacent Pa$\gamma$ line, are shown in Fig.~\ref{fig:Pa_gamma}.  The integrated fluxes of the two components, and their ratio, are listed in Table 4. 

\begin{figure}
\center
\plotone{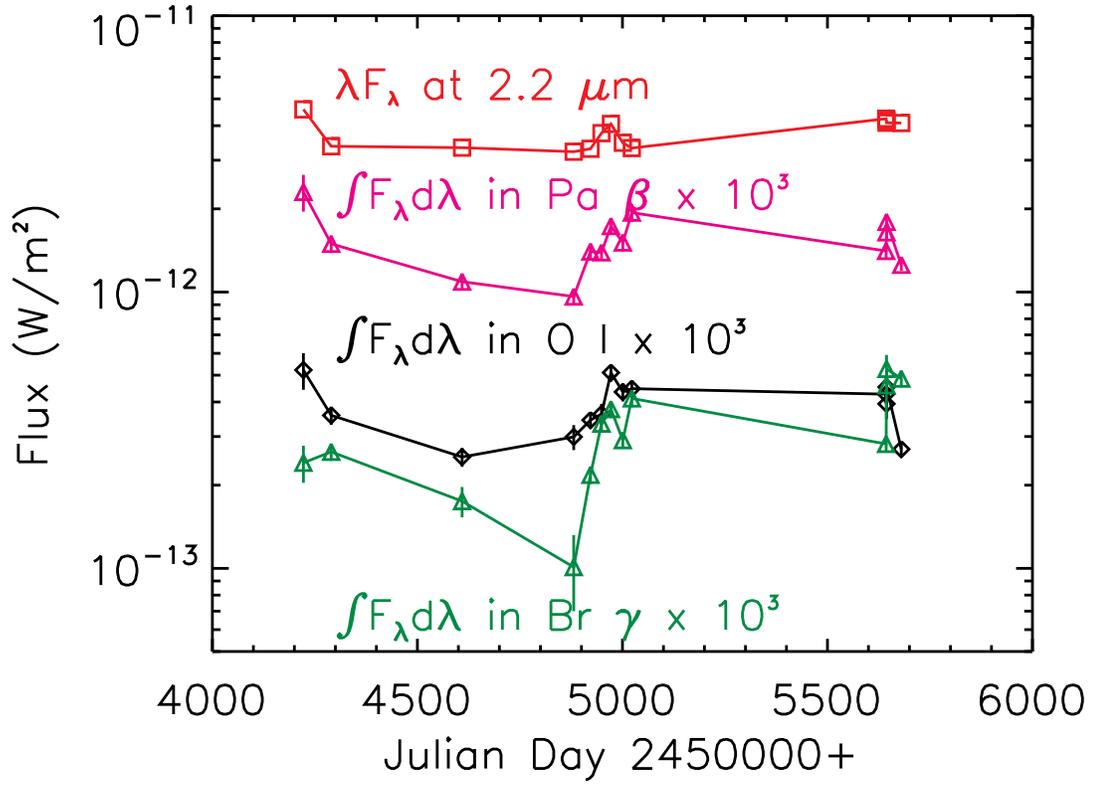}
\caption{The time variability  of Paschen $\beta$, Br $\gamma$, O I 0.8446 $\mu$m, and the K band for all 13 epochs . \label{fig:SpeX_lightcurve}}
\end{figure}

\begin{figure}
\center
\plotone{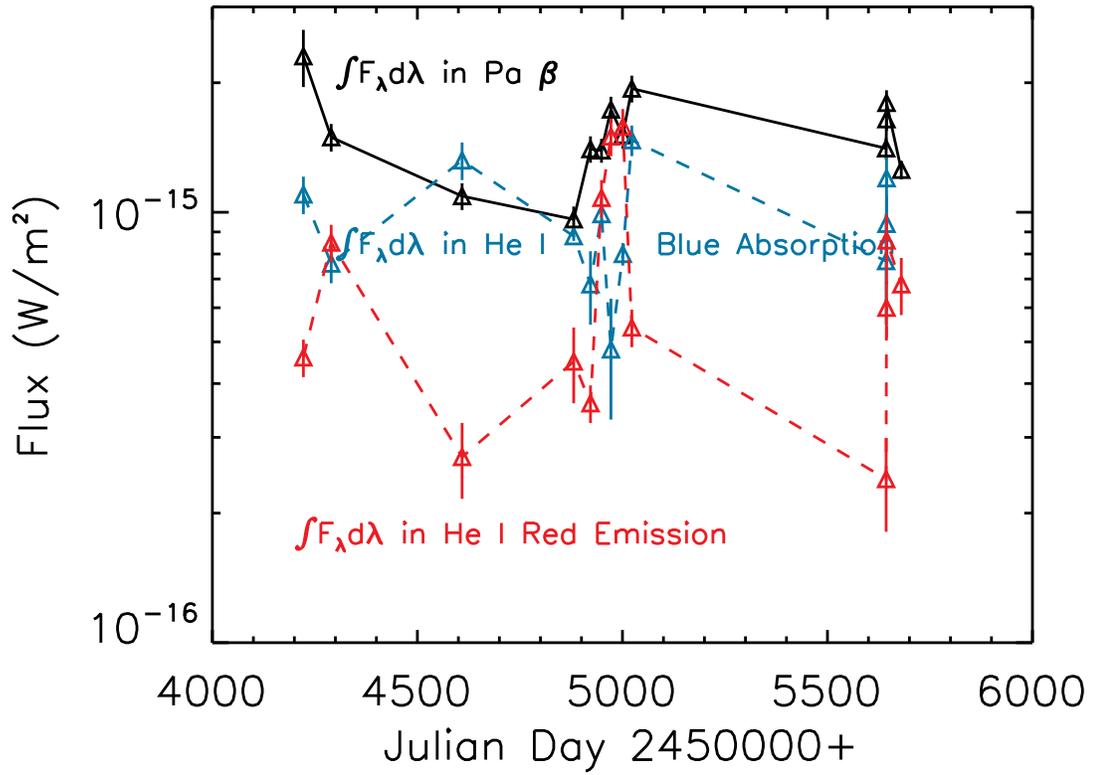}
\caption{The time variability of the emission and absorption components of He I 1.08 $\mu$m and the Paschen $\gamma$ emission lines for all 13 epochs. Here the integrated strength of the blue absorption component is a measure of the amount of energy \textit{removed} from the underlying continuum.   \label{fig:HeI_time}}
\end{figure}

Finally, Fig.~\ref{fig:SpeX_lightcurve} shows the light curves for the K band continuum, Pa$\beta$, Br$\alpha$, and O I line at 0.8446 $\mu$m lines. The resultant wavelength-integrated line strengths are given  in units of W m$^{-2}$ to match that of the continuum, which is in $\lambda$F$_{\lambda}$. Fig.~\ref{fig:HeI_time} compares the strengths of the He I net absorption and emission components with that of Pa $\gamma$.

\section{Character of the Variability and Mass Accretion Rates}

\subsection{Interrelations: Hydrogen, Helium, and Dust}

It is evident from examining Figs.~~\ref{fig:SpeX_lightcurve} and \ref{fig:HeI_time} that the dust, the hydrogen lines, and the He I line all behave differently.  Although the changing strength of the hydrogen and He I lines were first reported by  \citet{grady09}, the presence of only three data sets, and their limited cadence (2 months and one year) made it impossible to say much about the time scales of the variations, or to what degree the behavior of the hydrogen and helium lines were related. With the new observations and a shorter cadence it is apparent that although there is some ``jitter'', in the emission strength, the hydrogen line strengths seem to change on time scales of many months. By contrast, the He I emission and absorption can change dramatically in much shorter time scales. In 2011 March we observed SAO 206462 three times - once and 22 March and twice on 23 March, the latter two being separated in time by 2 hours.  While no significant change was seen between the two observations on March 23, significant changes occurred between March 22 and 23.

The origin and location of the line-emitting gas in disk systems is still puzzling, and a matter of controversy. The hydrogen lines are generally assumed to arise from the gas accretion onto the star, and this is precisely why their line luminosities are used to measure the accretion rates \citep{muzerolle98a,muzerolle98a,muzerolle01,calvet04}. The line strengths are generally well-correlated with other signs of accretion, such as veiling of the stellar spectrum by gas emission, ultraviolet line emission, etc. Spectrointerferometry by \citet{eisner10} indicates that, at least in the case of Br$\alpha$, the gas in younger disk systems is either in a magnetospherically driven accretion  and/or an outflow. However, similar observations of HD 104237 by \citet{tatulli07} suggested that its Br$\gamma$ emission line was  produced in a disk wind launched 0.2-0.5 AU from the star, and not the star itself.

For the He I line in such systems, \citet{edwards06} found that the broadest deepest portions of the absorption band were likely due to a stellar wind, while a more narrow low-velocity component arose from an inner disk wind. They suggested that some of the hydrogen lines might therefore also arise, in part, from a disk wind. In both wind sources, however, these should ultimately be driven by the accretion, since such absorption troughs are absent in non-accreting T Tauri stars. They noted that in some stars, the line is almost completely in absorption; in others almost completely in emission: ``the diversity of profile morphologies suggests that the nature of the winds are not identical in all of these accreting stars.'' More recently, \citet{kwan11} have suggested that both the H I and He I lines arise in clumpy winds. The rapid (1 day) time scale for the changes in the He I 1.083 \micron{} line are certainly consistent with an inhomogeneous flow.
 
In the case of SAO 206462, we can see that the He I strength of the emission versus the absorption is a matter of when the star was observed. The 2009 February data exhibit the weakest Pa $\gamma$ line strength as well as the weakest He I emission and absorption components. After the start of the outburst in 2009, the \textit{overall} strength of one or the other component of the He I line was larger, but which one dominated changed on time scales of a month or less. Without detailed modeling of the profile, which would require much higher spectral resolution, we cannot determine the details of the wind dynamics, but the simplest explanation of the changes observed is that the wind is not close to being spherically symmetric, and that the \textit{direction} of the wind with respect to the observer was changing significantly between epochs. On some epochs (2008 May 22 UT and 2011 March 22 UT) the absorption is strong and there is little if any net emission component visible. The line is totally dominated by gas flowing toward the observer. On other dates the emission was stronger than the absorption, but rarely totally dominating, indicating either a less-collimated flow, or one inclined further from the line of sight. Even at our low resolution, the absorption component in SAO 206462 would correspond with the broad stellar wind component discussed by Edwards et al.  

Do the hydrogen lines also possess a wind component?  The wind signature of the He I  arises due to the metastable nature of its lower energy state, a triplet state that cannot easily decay to the singlet ground state of the atom, so it behaves, in effect, like a ground electronic state.  By contrast, the Paschen lines have large transition rates to lower energy levels and are less likely to produce absorption unless H$\alpha$ has a high optical depth (leading to a significant population of the n = 2 energy level), even if they are in the wind.\footnote{The singlet He I line at 2.058 \micron{} is the counterpart to the triplet line at 1.083 \micron{}, leading to a permitted decay to the ground state, and should behave more like the Paschen lines. Unfortunately it is severely blended with a strong telluric band, making its use difficult for analysis.} Hence a wind contribution to the emission line strengths may be overlooked or underestimated. Much of the emission might not come directly from the accretion itself, but in the same wind that produces the He I line. Edwards et al. came to a similar conclusion. Thus the problem remains unresolved.

Compared to the observed changes in the gas line emission strengths, the dust continuum changes were only $\sim$30\% during the same span of time. The dust emission also began declining toward the end of the 2009 campaign, while the gas emission did not. The source of the changing dust emission might be related to events in the gas, but the correlation is not a strong one. As we showed in Fig.~\ref{fig:SpeX_lightcurve_JHK} all four continuum bands seemed to vary in unison during the 2009 campaign, with the exception of the March data, where the spectrum was steeper. This is best shown in Fig.~\ref{fig:3epochs} where the steepness of that data set is reflected in both the XD spectrum and the JHK photometry obtained at the same time. Thus the strength of the emission, as well as the inferred temperature, change with time.

\subsection{Mass Accretion Rates}

\citet{muzerolle98a} derived relationships between the strengths of the Pa$\beta$ and Br$\gamma$ lines and the total luminosity produced by the mass accretion onto the star for a mass range of 0.2-0.8 M$_{\sun}$. \citet{calvet04} extended the Br$\gamma$ calibration to 4 M$_{\sun}$. In Table 5 we show the net luminosities of the two lines, using the Muzerolle and Calvet calibrations for  Pa$\beta$ and Br$\gamma$, respectively. We used a photometric distance of $\sim$160 pc.\footnote{There is no parallax available from Hipparcos, but \citet{grady09} derived a distance to its common proper motion companion, SAO 206463, of 163 $\pm$ 3.8 pc, which is what we adopt for this paper. See Appendix A.} The uncertainties listed are those of the original measurements.\footnote{There are larger uncertainties in the calibrations themselves that can introduce an overall scaling to all the data.} These were then used to calculate the mass accretion rate assuming \.{M} $\approx$ L$_{acc}$R$_{star}$/GM$_{star}$.  Here we used R$_{star}$ = 2.3 R$_{\sun}$ and M$_{star}$ = 1.6 M$_{\sun}$. However, the radius of the star is not well-characterized.  \citet{grady09} suggested that v$_{eq}$=360$^{+80}_{-55}$ km s$^{-1}$. More recently, \citet{muller11}, using short cadence observations, suggested a rotational period of 3.9 hours. Coupled with their adopted radius of 1.4$\pm$0.25 R$_{\sun}$ gave v$_{eq}$=432$\pm$81km s$^{-1}$, while the break-up velocity was 480$\pm$60km s$^{-1}$. These equatorial velocities will cause the star to be significantly non-spherical.

The mass accretion rates derived from both bands varied between 1.5 - 4.2 x 10$^{-8}$ M$_{\sun}$ yr$^{-1}$ from Pa$\beta$, and 0.6-2.4 x 10$^{-8}$ M$_{\sun}$ yr$^{-1}$ from Br$\gamma$. Because the Br$\gamma$ calibration of Calvet et al.  includes stellar masses and line luminosities comparable to that of SAO 206462, it is likely to be the more accurate of the two, although there is considerable scatter in the data used to derive the calibration (see their Fig. 16).  \textit{Note that much of the scatter seen in the data used in the calibrations of emission line strength and other accretion rate indicators may be the result of real variability when the various data sets used are not obtained close enough in time} (see Appendix B and Fig.~\ref{fig:correlation}). Regardless, the line luminosities, and hence the accretion rates, varied by factors of 2-4 over a time span of 6 months.

By comparison, the change in strength of the O I line was less than either that of Pa$\beta$ or Br$\gamma$.  This could either be due to the difficulty of measuring such a weak line, or its sensitivity to different excitation mechanisms. While the H I lines are due to recombination, the O I 0.8446 $\mu$m line is usually presumed to be dominated by Bowen fluorescence from hydrogen Ly $\beta$ excitation, although \citet{rudy89,rudy91} mention two other possible contributions to its production in circumstellar environments - recombination and continuum fluorescence by the star.  A possible test of whether Ly$\beta$ pumping dominates would be to measure the O I line at 1.1287 $\mu$m, which would be produced in equal numbers of photons as the 0.8446 $\mu$m line, but that transition sits in a wavelength region of significant telluric absorption.

\section{Discussion}

\citet{sitko08} discussed a number of mechanisms for the production of variability in the inner disk of of the Herbig Ae star HD 163296, including thermal instabilities, magnetorotational instabilities (MRI), X-wind effects, and planetary perturbations. \citet{flaherty11} have also illustrated a wide range of possibilities for the source of the variability observed in the transitional disk system of LRLL 31. Regardless of the mechanism(s) involved, the relatively short time scales involved in the continuum variability of SAO 206462, less than a year, implicate the innermost regions of the dust belt or the star itself. In SAO 206462, the continuum and lines also do not vary in precisely the same manner. An examination of the light curves show that both the dust emission and the gas emission seemed to be increasing together at the beginning of the 2009 observing campaign, but the dust emission subsided after a few months, while the gas emission continued to increase. It is also evident that the behavior of the He I 1.083 $\mu$m line is much more complex than that which was observed for the Brackett and Paschen lines, with significant changes  occurring on time scales of a day or less. 

Any attempt to understand the source of the variations in the gas and dust emission requires some knowledge of the structure of the dust belt, and its relation to the star. \citet{brown07} modeled the SED of SAO 206462 using a grain model with a steep power law size distribution and small and large size limits that were larger than those thought to apply to interstellar values, consistent with the lack of a significant 10 $\mu$m silicate band, as well as what would be expected from grain growth. For their model the locations of the inner and outer radii of the warm inner dust belt were 0.18 and 0.45 AU, respectively. The actual distances derived by SED-fitting will be sensitive to the grain sizes and compositions of the model - well-known the degeneracy problem with such models where the grain temperatures are sensitive to both their optical properties and distance from the star. In addition, the data sets used by \citet{brown07} were not simultaneous, and in fact they do not overlap very well at the very wavelengths where the contribution of the light of the dust belt to the SED is greatest. Using a different grain model, \citet{grady09} derived distances of 0.08-0.24 AU (for an epoch with a steeper near-IR SED) and 0.14-0.31 AU (for an epoch with a flatter near-IR SED). In these cases, the 0.8-5 $\mu$m data were simultaneous, but the fits to the data longward of 3 $\mu$m for the steeper SED data were not particularly good.

Interferometry of the inner dust belt by \citet{fedele08} indicated an outer radius of 12.8 $\pm$ 1.4 mas which, for our adopted distance of 160 pc, corresponds to 2.0 $\pm$ 0.2 AU. The inner radius was not resolved\footnote{\citet{fedele08} formally report a value of 0.35$\pm$1.79 mas} but a 3-$\sigma$ upper limit would be 5.4 mas corresponding to 0.86 AU. Thus the outer edge of the dust torus  is likely 2 AU or smaller, depending on the true distance to SAO 206462), but a more precise value is not yet available. 
 
The lack of resolution of the inner edge by the interferometry may signal the presence of some material as close as 0.1 AU. Material inside the usual dust ``rim'' in the disk of the Herbig Ae star HD 163296 detected interferometrically has been reported by \citet{ajay08a} and \citet{benisty10}. Tannirkulam et al. suggested that the detected emission was due to gas located interior to the dust rim. As previously noted, Benisty et al. favored super-refractory dust grains, such as iron, graphite, and corundum, with graphite having a greater likelihood of surviving the extreme temperatures required by their model. Such grains may be present interior to the sublimation radius of silicate grains in main pre-main sequence (PMS) disk systems. For SAO 206462,  \citet{grady09} had suggested the possibility of carbon-rich grains in the inner torus of SAO 2966462. Such grains could easily survive close enough to the star to prevent the inner edge from being resolved.
 
\citet{fedele08} also determined the location of one gas component, the [O I] gas, from its 6300 \AA \  line profile. Roughly two-thirds of the gas was coincident with the location of the inner dust bely, inside 2 AU, with a more distant local minimum between 2 and 7 AU, and increasing abundance again at greater distances. Their assumed inclination was 60$^{\circ}$. However, \citet{dent05} derive an inclination of the CO gas as 11$\pm$2$^{\circ}$, while  \citet{pontoppidan08} get 14$\pm$3$^{\circ}$. For these smaller values of the inclination, the gas must be 2-5 times closer to the star.\footnote{Both \citet{grady09} and \citet{muller11} discuss the possibility of different inclinations for the inner dust belt and the outer disk. The latter authors suggest that the maximally-rotating star is losing mass to the outer disk equatorially in a \textit{decretion disk} reminiscent of classical Be stars \citep{wisniewski10}, while the tilted inner dust belt is feeding the accretion inward to the star.} Regardless, this places the minimum of the surface density of the gas very close to, or just outside of, the outer edge of the inner dust belt.

The inclination may also affect how we would view the stellar wind. If the star has an inclinations less than 20$^{\circ}$, then magnetically-funneled accretion columns and outflow jets would likely be pointed toward the observer if they were predominantly located near the polar axis. A slightly off-axis outflow would be modulated by the stellar rotation. For the rotational period of the star derived by \citet{muller11} this would be measured in \textit{hours}. The two observations we obtained on 23 March 2011, separated by one-half of this period, showed no significant changes, but more short-cadence observations are required  to adequately investigate this possibility.

What is the vertical extent of the dust belt? The SED of SAO 206462 \citep{grady09} requires that the belt intercepts $\sim$25\% of the stellar luminosity. If it is optically thick, it must subtend a solid angle of $\sim$$\pi$ ster as seen from the star (these numbers are somewhat uncertain as the belt likely ``sees'' the star equator-on, while the observer is seeing it more pole-on). The half-opening angle (mid-plane to its ``top'') would need to be $\sim$15$^{\circ}$ if it intercepts all the light incident on it, or even higher if it is optically thin. The minimum value that the optical depth could have is $\sim$0.3, at which point it is not a belt at all, but a halo. \citet{krijt11} have discussed how a dynamically excited cloud of planetesimals in could lead to the production of such a halo. For a low-inclination system a spherical shell might appear interfreometrically like a disk with material inside its inner ``rim''; the two geometries might be distinguished by the presence or absence of shadowing on the inner rim of the outer disk.

\subsection{Thermal Instabilities and the Inner Dust Ring}

Because SAO 206462 is still actively accreting, it may be subject to instabilities believed to be present in such disk systems. Any mechanism which would lead to an increase in the heating of the disk will require time for it to dissipate. Here we assume that the inner gas is mixed at least in part with the inner dust belt. Following the discussion in \citet{sitko08} we use the relationships of \citet{ss73} and \citet{pringle81} to investigate some of the time scales relevant for changes in a standard ``$\alpha$ disk''. Within that framework, the shortest time scale is  the ``dynamic'' time scale for a parcel of material (such as a hot spot at the disk rim) to move one radian it its orbit and is given by

$t_{\phi}=\frac{R}{v_{\phi}}=\Omega^{-1}$

\noindent where R is the radius outward in the disk, $v_{\phi}$ is the orbital velocity, and $\Omega$ is the angular velocity, so that the orbital period is 2$\pi$$ t_{\phi}$. For R in AU, $t_{\phi}$ in years, and the stellar mass M in solar masses, Keplerian orbits will have

$t_{\phi}=\frac{1}{2\pi} \frac{R^{3/2}}{M^{1/2}}$.

The thermal time scale, the time it takes for the disk to radiate a significant fraction of its internal heat,  is then 

$t_{th} \sim \frac{1}{\alpha} t_{\phi}$  = $\frac{1}{2 \pi \alpha}$ P

\noindent where P is the orbital period and $\alpha$ is the viscocity parameter. At a distance of 0.5 - 2 AU the orbital periods are $\sim$0.2-2.2 years, and the thermal time scale  would be

$t_{th} \sim \frac{0.04}{\alpha}-\frac{0.44}{\alpha}$ years. 

If $\alpha \sim$  0.01-0.001, at the location of the outer edge of the dust belt, the thermal time scale is over a half decade, as opposed to a few months that we observe the variations in the dust to have, and this mechanism can be ruled out. We caution that these time scales are derived for disks where viscous heating dominates, whereas in transitional disks the stellar irradiation will make a much larger contribution to the disk heating. In the case of SAO 206462, the accretion luminosities we derive are $\sim$0.5 L$_{\sun}$, while the stellar luminosity is 4-8 L$_{\sun}$ for a pole-on view, and likely at least half-that as seen by the dust belt. \citet{frank02} suggest that the stellar irradiation of the disk will dominate at all radii if L$_{star}\ge$7.5L$_{acc}$(1-$\beta$)$^{-1}$, where $\beta$ is the grain albedo. For $\beta$$\sim$0.5, this condition is L$_{star}\ge$15L$_{acc}$, very close to the relative values for SAO 206462. So in the case of SAO 206462, the heating of the disk is may dominate, which would tend to even out changes in heat content by this mechanism.

\subsection{Magnetorotational Instabilities (MRI) and the Inner Dust Ring}

In the simulations of \citet{suzuki09} the MRI begins to develop after a few rotations, but becomes quasi-steady-state after about 200 rotations. But mass fluxes near the surface region, which can lead to disk winds, are highly time dependent and have quasiperiodic cycles of $\sim$ 5 - 10 rotations. At the location of the dust belt in SAO 206462  this corresponds to years, and would seem to be too long to be the source of changes in this structure. Thus evidence that this mechanism is operating between 0.5 and 2 AU in an inner dust belt is also weak.

\subsection{Instabilities in the Inner \textit{Gas} Disk}

One of the problems with explaining the observed changes as arising from mechanisms operating near the interferometrically-resolved outer region of the dust belt is that they take a long time to develop, and it is unclear how they manage to make such a large change in the gas accretion onto the star that powers the emission line flux. While the thermal instabilities and the MRI would operate on time scales of a decade or more in that region, changes will be more rapid closer to the star, where the disk might be largely dust-free if super-refractory grains are not present. At a distance of 0.2 AU, the MRI time scale is closer to 3 months, consistent with the development of enhanced line emission seen in SAO 206462. The standard $\alpha$ disk thermal instabilities are probably irrelevant here - the time scales are still too long, and it is unlikely that significant changes in the thermal content in the disk could occur if the MRI is also operating and has reached quasi-steady-state. The variations might occur more rapidly when they are dominated by surface, rather than volume, effects. \citet{muller11} find a significant periodicity in the strength of the H$\alpha$ line of 5.77 days, corresponding to the orbital period at $\sim$0.25 AU. The periodic changes were likely the cause of the ``jitter'' in the strength of Pa$\beta$ and Br$\gamma$ (Fig.~~\ref{fig:SpeX_lightcurve}). \footnote{The most extreme point in the equivalent width measurements in \citet{muller11} was obtained 90 days prior to the rest, suggesting larger changes occur on longer time scales, similar to those presented here.}

\subsection{Planetary Perturbations and the Inner Dust Ring} 

If the gap in the dust distribution (inner belt and outer disk) and the local minimum in the [O I] gas are due to the gravitational effects of a massive planet, it would interact with the inner belt in a periodic fashion if its orbit were either eccentric or highly inclined.  This seems to be the case for many exoplanets orbiting hot stars  \citep{winn10}. A planet whose periastron is close to 2 AU would have an orbital period of slightly longer than 2 years, and would likely interact with the disk on roughly that time scale, inducing off-center orbits or disk warpage. Future observations with more complete time coverage could test this hypothesis.

If on the other hand, the spacing between the inner and outer dust structures is not a gap opened by a planet, but rather, if the inner ring is a dusty structure regenerated by collisions \footnote{This is not the standard older gas-poor  debris disk at at a large distance from the star, but a small one restored through hypervelocity impacts closer to the star, as suggested by \citet{lisse09} for HD 172555.}, then a planet need not occupy the gap between the inner dust belt and the outer disk. It could instead reside closer to the star than the dust belt. This would cause it to have a shorter period, but probably not less than 6 months, or the inner dust ring would be in a state of almost constant excitation.  This means that the planet would probably need to orbit more than 1 AU from the star. This is within the range of the location of the belt itself and suggests the possibility of the inner dust ring being located at a 1:1 gravitational resonance with a planet. Such structures and their observable effects are being investigated for more distant regions in ptotoplanetary and debris disk systems \citep{fouchet10,stark08,hahn10} but nothing prevents them from being located closer to the star, as the dust belt associated with the Earth \citep{reach10} attests to.

\subsection{X-winds}

The X-Wind models discussed by \citet{shu94} describe how the stellar magnetic field is connected to that of the differentially-rotating gas disk. It has a natural advantage in providing a mechanism for producing collimated outflows, and automatically includes the possibility that reconnection events and other variations in the magnetic field could alter both the outflow rate and structure in the ionized gas. As	 previously noted, the rotational period of the star is likely measured in hours. If the magnetic connection between the disk and star has been strong, then it should have magnetically braked the rotation of the star well below its observed rotational speed - the X-wind model was developed in part to \textit{explain why} classical T Tauri stars were relatively \textit{slow} rotators, despite the expected spin-up due to the angular momentum of disk material being deposited onto the star. 

For SAO 206462, the longer-term modulation (unlike the shorter-term scatter discussed in \S2.5) in the visible-wavelength photometry of the photospheric flux is not unlike that seen in later-type stars and attributed to star spots. This would imply surface magnetic field activity. This would be consistent with the observed X-ray luminosity of 3.6 x 10$^{28}$ erg s$^{-1}$ (for d=160 pc) observed for this star (Alex Brown, private communication). 

\subsection{Summary}

Of the mechanisms discussed here  for the variability in the inner dust belt, forcing by planetary perturbations has the right time scale for the changes observed to date. Thermal instabilities would tend to require longer time scales, and are less likely to operate where the heating of the disk is being controlled to a significant degree by the star. Similarly, the MRI will likely act on time scales longer than those observed here, unless it is acting primarily on the innermost gaseous portion of the disk. This might help drive the changes seen in the line intensity, and temporarily affect the nearby dust band.  With a rotational period less than four hours, it is unlikely that the magnetic field of the star is connected to the inner disk of accreting gas, a requirement of the X-wind model. One scenario yet to be explored is tied to the rapid rotation of the star. If it in fact is shedding gas from its equator, this gas may interact with the dust belt in ways that will need to be investigated.

\section{Conclusions}

Until recently, most studies of gas accretion and thermal dust emission operated in an atmosphere of ``steady state'' wherein any description of disk characteristics, accretion rates, and other relevant parameters were expected to be accurate descriptions of the given object(s) under study. But the recent detection of the variability of gas emission lines \citep{sitko08,grady09,eisner10}  and dust emission \citep{woodward04,sitko08,grady09,muzerolle09,bary09,rebull10,espaillat11} on relatively short time scales indicates that this is not the case. In at least one case, HD 163296, interferometry indicated a \textit{structural} change on time scales of a year \citep{ajay08a,ajay08b}. Similar interferometric monitoring of these systems over a range of time scales will be required to determine precisely how the structure of the inner disk regions are changing.

In SAO 206462 the mass accretion rate derived from the emission lines is a few times 10$^{-8}$M$_{\sun}$y$^{-1}$, and can change by a factor of more than 2 over the course of a few months. It is unclear how much of the observed emission is coming directly from the accretion onto the star, and how much from an outward flowing wind. The He I line at 1.083 $\mu$m exhibits the profile of a classic stellar wind, and not of infalling gas - none of the 13 epochs show an inverse P Cygni line profile indicative of infalling gas - and by inference many of the H line emission used to measure the accretion rate must, in fact, be produced in part by outflowing gas as well, but the gas is not optically thick enough to produce an observable blue-shifted absorption component. Spectrointerferometry of the emission lines in PMS stars suggests that most of the emission in lines such as Br$\gamma$ is consistent with polar inflows or outflows \citep{eisner10}, but these observations cannot currently distinguish between the two.

The degree (factors of 2 or more) and and time scale (months) of the changes in the observed line intensities suggest that a significant amount of the scatter in the existing  calibrations of accretion rate versus line luminosity may be the result of using non-simultaneous data sets. The near-IR spectra used by \citet{calvet04} for the calibration of gas accretion rates were obtained in January 1998, a full two years prior to their \textit{Hubble} spectra and visible-wavelength photometry (February and September 2000), and over a year before much of their visible-wavelength spectroscopy (September 1999).  These epochs are easily different enough that they are not measuring the system at the same state. A tighter calibration might be possible with coordinated nearly-simultaneous observations. We discuss this further in  Appendix B.

Gapped pre-transitional disks such as the one surrounding SAO 206462 are generally considered signposts of planet formation, and do not suffer the ambiguity of the fully open inner disks of their more evolved transitional disk counterparts, where mechanisms other than planet formation may be responsible for the inner disk clearing. We have detected significant changes in the gas and dust emission in this system whose origin is still unclear. Typical disk instabilities (thermal, MRI) would seem to operate on time scales longer than what is observed in the warm inner dust belt of SAO 206462, and are probably excluded as the source of its variability, at least in the outer (0.5-2 AU) portion of the belt. There, planetary interactions operating on orbital time scales are consistent with those actually observed.  Furthermore, if gapped disks require the presence of planets to sweep up disk material in order to produce the gap, \textit{the planets must, by definition, be interacting with the disk}, and these interactions may be responsible for much of the behavior observed. To interact with the inner dust ring of SAO 206462, the planet would likely have to have an orbital period of 1-3 years.  Disk instabilities operating on the \textit{gas} much closer to the star might help initiate accretion or wind events on short time scales, but how these are related to the emission of the dust belt remains to be investigated.

Significant progress in solving the origin of the variations on the dust and gas emission in SAO 206462 and other transitional, pre-transitional and other young disk systems, can nevertheless be made, with carefully-designed ``experiments''. These include:

A. Observations of the thermal emission of the dust on time scales commensurate with the orbital periods of planets that may be sculpting their boundaries. This strategy is not unlike that used to find the planet interacting with the cold, off-center debris disk in Fomalhaut \citep{kalas09}. In the case of SAO 206462, this will require carefully monitoring over time spans of a number of years to confidently detect periodic changes in the outer edge of the inner dust belt. Planets closer than this may also be present, so well-sampled shorter cadences need to be included as well.

B.  Observations of the line emission on time scales shorter that the rotational periods of the stars. If significant He I line profile changes are detected on time scales comparable to the rotational periods of the star, then much of the changes observed might arise from changes in the orientation of a collimated wind with respect to the observer. For early-type spectral classes that have not undergone significant rotational braking, this means time cadences of hours, or sufficient data that can be adequately phased to the rotation period to look for these changes. Our single pair of observations with this time spacing are not sufficient to completely address this problem.

C.  Repeated near-IR interferometry with cadences that can sample the gas and dust on time scales relevant to changes that may occur in different regions of the disk.  This will provide the clearest information on the source of the variability, provided these regions can be spatially resolved. However, such observations are very expensive in terms of their use of available resources, as multiple telescopes will need to be used repeatedly.

\acknowledgments

The authors would like to thank Bill Vacca, Mike Cushing, and John Rayner for many useful discussions on the use of the SpeX instrument and the Spextool processing package. We also thank the entire REM team for their assistance with the scheduling and execution of those observations. Thanks also to Alex Brown for the use of the unpublished X-ray luminosity information. More thanks to Nuria Calvet, David Wilner, John Monnier for many useful discussions about this star and related objects.  This work was supported by  NASA ADP grants  NNH06CC28C \& NNX09AC73G, \textit{Hubble Space Telescope} grants  HST-GO-10764 \& HST-GO-10864, and the IR\&D program at The Aerospace Corporation.

{\it Facilities:} \facility{IRTF (SpeX,BASS)}

\appendix

\section{Adopted Stellar Parameters for SAO 206462}

Lacking an accurate parallax value from Hipparcos, the distance to SAO 206462 is uncertain. \citet{grady09} estimate the distance to its A0V companion, SAO 206463, as 167 pc  and the reddening as E(B-V)=0.078 mag, while \citet{vanboekel05}  suggests 140 pc if it is a member of the Sco OB2-3 association. These yield Log L $\sim$ 0.96 L$_{\sun}$ and $\sim$ 0.81 L$_{\sun}$, respectively. The spectral type of F4Ve \citep{dunkin97} has been adopted here. With a reddening-corrected (B-V)$_{0}$=0.42,  T$_{eff}$ = 6600 K using the T$_{eff}$ vs. color calibration of \citep{ramirez05}, assuming main sequence and solar metallicity; but it is 6500 K if it is closer to F4 III.

Using the zero-age main sequence fitting of NGC 2516 by \citet{an07}  (which includes the (B-V) of SAO 206462), a true main sequence star with (B-V)$_{0}$ = 0.42 would have, after correcting for the reddening,  an absolute visual magnitude of M$_{v}$ = 3.27 mag for (m-M)$_{0}$ = 8.03, their derived value, while M$_{v}$ = 3.60 for (m-M)$_{0}$ = 7.70, based on the Hipparcos distance derived by \citet{robichon99}. These give Log L = 0.60 and 0.46, respectively. For the ``long'' distance of 167 pc, this means SAO 206462 is  0.92 or 1.25 mag above the main sequence  for L$_{MS}$ = 0.60 or 0.46, respectively. For the short 140 pc distance it is 0.53 - 0.86 mag above the MS.

These values are generally consistent with the star being a rapid rotator seen nearly pole-on \citep{grady09}, similar to Vega. For T$_{eff}$ = 6600 K and log L = 0.96 L$_{\sun}$ we get R = 2.3 R$_{\sun}$; for log L = 0.81L$_{\sun}$ it is 2.0R$_{\sun}$.

Using these luminosities and T$_{eff}$ = 6600 K and the evolution tracks of \citet{dantona94} we get M$\sim$1.6 M$_{\sun}$ (167 pc) and 1.5 M$_{\sun}$ (140 pc). In an independent analysis \citet{muller11} adopt d=142$\pm$27 pc, T$_{eff}$=6810$\pm$10 K, R$_{star}$ = 1.4$\pm$0.25 R$_{\sun}$ and M$_{star}$ = 1.7$^{+0.2}_{-0.1}$ M$_{\sun}$.

\section{Use of Non-Simultaneous Data Sets}

Deriving correlations between non-simultaneous data sets in variable sources will lead to an increase in scatter, compared to those obtained at the same time. Because these relationships are often derived from a wide variety of instruments, contemporaneous scheduling is sometimes difficult to achieve. In this paper, we utilized the calibrations of the Pa$\beta$ and Br$\gamma$ lines of \citet{muzerolle98b} and \citet{calvet04} which were observed at a different epoch than the other accretion indicators.

\begin{figure}
\center
\plotone{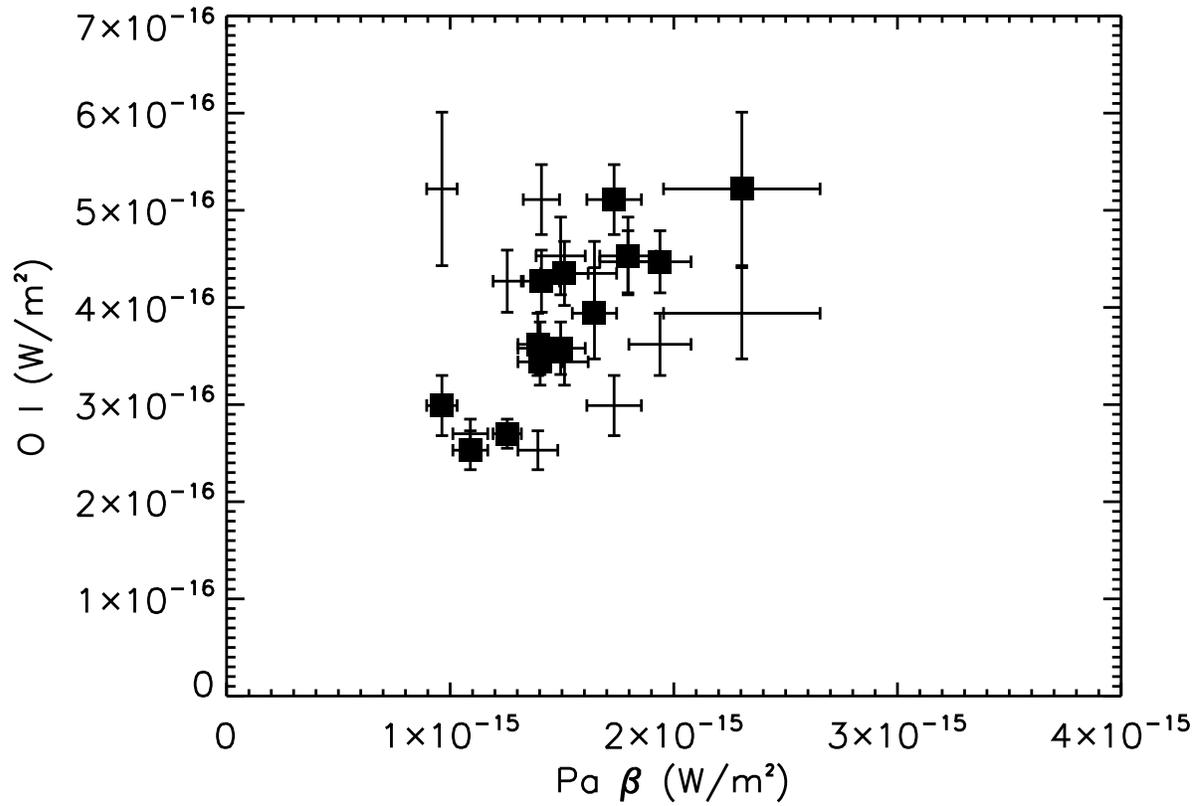}
\caption{The relative strengths of Pa$\beta$ and O I $\lambda$8446. The filled symbols represent the actual observations, with matching dates of observation. The data shown only as error bars represent the same data set ``rotated'' by three epochs, to illustrate in a simple manner the result of using two non-simultaneous data sets.  \label{fig:correlation}}
\end{figure}

In  Fig.~~\ref{fig:correlation} we show the effects using the Pa$\beta$ and O I line strengths from our data sets. Here we show the data displayed twice: in one case, the  O I line strengths are plotted versus the Pa$\beta$ strengths obtained on the same night. In the other set, we simply rotated the timing of the O I observations by three epochs, in order to create a simulation of the effects of using data sets separated in time by time spans longer than the time scale for variability of the lines. Thus the true intrinsic scatter in the calibrations of the luminosity of these two lines  with the accretion rate derived from other indicators is likely better than the original figures in \citet{muzerolle98b} and \citet{calvet04} would indicate.

\begin{deluxetable}{lcccc}
\tablecolumns{5}
\tablewidth{0pc}
\tablecaption{SpeX Observations}
\tablehead{
\colhead{UT Date} & \colhead{Mode} & \colhead{SAO 206462 Airmass} &  \colhead{Calibration Star Airmass} & \colhead{Seeing (arcsec)} }

\startdata
 2007.05.01 & SXD & 1.84 & 1.75 &  0.7 \\
                      & LXD & 1.84 & 1.75 &       \\ 
 &  & &  &   \\
 2007.07.08 & SXD\tablenotemark{a} &  1.95 & 1.75 & 0.5  \\
 &  & &  &   \\
2008.05.22 & SXD & 2.31 & 1.95 \& 2.77 & 1.0  \\
                     & LXD & 2.09 & 2.05 - 2.12 &  \\ 
                     & Prism & 1.87 & 1.93 &  \\ 

 &  & &  &   \\                     
2009.02.18 & SXD & 2.13 & 2.06  &  ---\tablenotemark{b} \\ 
                     & LXD & 1.86 - 2.01 & 1.92  &   \\ 
                     & Prism & 1.84 & 1.76 &   \\
&  & &  &   \\       
2009.03.31 & SXD & 1.93  & 1.75 & $\sim$2  \\
                     & LXD & 1.84  & 1.76 &  \\ 
                     & JHK & 1.86  & 1.74 &  \\
&  & &  &   \\ 
2009.04.27 & SXD & 1.83  & 1.82 & 0.5   \\
                     & LXD & 1.89  & 1.83 - 1.86 &   \\
                     & Prism & 1.83  & 1.81 &   \\
&  & &  &   \\ 
2009.05.20 & SXD & 1.85 & 1.89 & 0.8   \\
                     & LXD & 1.98 & 1.93-1.97 &    \\ 
                     & JHK & 2.30 - 2.34 & 2.37 - 2.39 &    \\
&  & &  &   \\ 
2009.06.18 & SXD & 2.08 & 2.38 & 1.0 \\
                     & LXD & 2.54 \& 2.75 & 2.45 - 2.65 &  \\ 
                     & K & 2.50 & 2.71 - 2.74 &  \\
&  & &  &   \\ 
2009.07.10 & SXD & 1.94 & 1.82 & 0.9  \\
                     & LXD & 1.88 & 1.85-1.88 &   \\ 
&  & &  &   \\ 
2011.03.22 & SXD & 1.84 & 1.74 & 0.7 \\     
                      & LXD  & 1.96 & 1.85  & \\            
                      & Prism & 1.91 & 1.75 & \\
&  & &  &   \\ 
2011.03.23 \tablenotemark{c} & SXD & 2.00 &  2.01 &  1.1 \\
                     & LXD  &  1.92 & 1.92  & \\
                      & Prism & 1.86 &  1.82 & \\  
                      & SXD &  1.83 & 1.88 &  \\
&  & &  &   \\ 
2011.04.28 & SXD & 2.05 & 2.06 & 0.5 \\
                      & LXD & 2.30 & 2.33 &  \\
                     & Prism & 1.92 & 1.97 &  \\

\enddata
\tablenotetext{a}{SXD only due to cirrus.}
\tablenotetext{b}{No record of the seeing was made.}
\tablenotetext{c}{Two separate SXD observations beginning at 1214 UT and 1344 UT, respectively.}
\end{deluxetable}

\begin{deluxetable}{lccc}
\tablecolumns{4}
\tablewidth{0pc}
\tablecaption{BASS Observations}
\tablehead{
\colhead{UT Date}  & \colhead{SAO 206462 Airmass} &  \colhead{Calibration Star Airmass} & \colhead{Calibration Star}}

\startdata
2007.07.09 & 1.87 - 1.93 & 1.85 - 1.90 & Vega \\
2009.07.14 & 1.84 - 1.88 & 1.70 - 1.75 & Vega \\

\enddata
\end{deluxetable}

\clearpage

\begin{deluxetable}{cccccccccc}
\tablecolumns{10}
\tablewidth{0pc}
\tabletypesize{\scriptsize}
\tablecaption{Line and Continuum Fluxes\tablenotemark{a}}
\tablehead{
\colhead{Date (UT)} & \colhead{O I}  & \colhead{Pa $\epsilon$} &  \colhead{Pa $\delta$} & \colhead{Pa $\gamma$}  & \colhead{Pa $\beta$}  & \colhead{Br $\gamma$ }  & \colhead{Br $\alpha$}  & \colhead{K} }

\startdata
070501 & 5.22(0.79) &  4.92(0.75) & 14.94(2.75) & 16.41(2.55) &  23.04(3.50) &  2.41(0.37) & 5.81(0.88) & 4.58(0.69) \\
070708 &3.58(0.27) &  4.83(0.37) &  8.69(0.79) &  11.19(0.95) & 14.94(1.10) & 2.64(0.19) & ---\tablenotemark{\ b}  & 3.37(0.24)  \\
080522 & 2.53(0.20) &  5.40(0.39) & 7.53(0.74) &  9.34(0.71) & 10.90(0.78) & 1.75(0.22) &  3.97(0.30)  & 3.33(0.23)  \\
090218 &  2.99(0.31) & 3.84(0.33) & 7.89(0.80) & 6.74(0.52) & 9.63(0.68) & 1.01(0.31) &  3.19(0.23) & 3.22(0.22) \\
090331 &  3.44(0.24) & 4.61(0.44) & 7.89(0.73) & 7.72(0.54) & 14.02(0.99) & 2.18(0.16) &  4.38(0.31) & 3.30(0.23) \\
090427 &  3.62(0.32)  & 5.65(0.40) & 8.53(0.65) & 10.91(0.89) &  13.92(0.89) & 3.34(0.24) & 3.98(0.29) & 3.76(0.26) \\ 
090520 &  5.11(0.36) & 8.45(0.59) &  12.60(1.07) & 13.04(0.96) & 17.33(1.22) & 3.77(0.27) & 5.26(0.37) & 4.06(0.28) \\
090618 & 4.35(0.33) & 7.42(0.53) & 10.34(0.93) & 11.70(0.89) & 15.11(1.06) & 2.91(0.20) & 3.47(0.33) & 3.47(0.24) \\
090710  & 4.47(0.32) & 6.69(0.50) & 11.57(1.25) &  14.77(1.18) & 19.38(1.39) & 4.12(0.29) & 5.05(0.40) & 3.32(0.23) \\
110322 & 4.27(0.32) & 5.93(0.31) & 8.11(0.80) & 10.09(0.77) & 14.08(0.81) & 2.82(0.16) & 5.52(0.37) & 4.24(0.30) \\
110323a & 3.94(0.47) & 7.01(0.40) & 9.88(0.85) & 11.62(0.71) & 16.45(0.99) & 4.60(0.25) & 3.94(0.23) & 4.18(0.29) \\
110323b & 4.53(0.40) & 8.76(0.69) & 12.85(1.54) & 13.82(1.27) & 17.95(1.26) & 5.27(0.65) & 3.87(0.23) & 4.10(0.29) \\
110428 & 2.70(0.15) & 4.70(0.52) & 7.96(0.78) & 9.76(0.66) & 12.55(0.63) & 4.86(0.34) & 4.25(0.36) & 4.09(0.29) \\
\enddata
\tablenotetext{a}{ ~Line fluxes are $\int$ F$_{\lambda}$d$\lambda$ in units of 10$^{-16}$ W m$^{-2}$ while the K-band flux is $\lambda$F$_{\lambda}$in units of 10$^{-12}$ W m$^{-2}$. Uncertainties are given in parentheses.}
\tablenotetext{b}{~No LXD spectrum was obtained.}

\end{deluxetable}

\clearpage

\begin{deluxetable}{lccc}
\tablecolumns{4}
\tablewidth{0pc}
\tablecaption{Emission and Absorption Strengths of the He I 1.083 $\mu$m Line\tablenotemark{a}}
\tablehead{
\colhead{Date (UT)} & \colhead{Blue Absorption}  & \colhead{Red Emission} &  \colhead{ Ratio Blue/Red} }

\startdata
070501 & 10.0 (1.1) & 4.6 (0.5) & 2.4 (0.3)  \\
070708 &  7.6 (0.8) & 8.5 (0.9) & 0.9 (0.1) \\
080522 &   13.2 (1.3) & 2.7 (0.5) & 4.9 (1.1) \\
090218 &   1.5 (0.2) & 4.5 (0.9) & 0.3 (0.1) \\
090331 &  8.8 (1.3) & 3.6 ( 0.4) & 2.4 (0.4) \\
090427 &  6.8 (0.7) & 10.8 (1.1) & 0.6 (0.1)  \\ 
090520 &   9.9 (1.5) & 15.0 ( 1.5) & 0.7 (0.1) \\
090618 & 4.8 (0.5) &  15.8 (1.6) & 0.3 (0.04) \\ 
090710  & 8.0 (1.2) & 5.4 (0.5) & 1.5 (0.3)  \\
110322 & 14.7(2.2) & 2.4(0.6) & 6.20(1.81) \\
110323a & 7.7(1.5) & 6.0(0.9) & 1.29(0.32) \\
110323b & 9.4(1.4) & 8.6(1.3) & 1.10(0.23) \\
110428 & 12.0(1.8) & 6.8(1.0) & 1.75(0.37) \\

\enddata
\tablenotetext{a}{ ~Line fluxes are $\int$ F$_{\lambda}$d$\lambda$ in units of 10$^{-16}$ W m$^{-2}$. Uncertainties are given in parentheses.}

\end{deluxetable}

\clearpage

\begin{deluxetable}{ccccccc}
\tablecolumns{7}
\tablewidth{0pc}
\tabletypesize{\scriptsize}

\tablecaption{Br $\gamma$ and Pa $\beta$ Luminosities\tablenotemark{a} and Mass Accretion Rates\tablenotemark{b} in SAO 206462}
\tablehead{
\colhead{Date (UT)} & \colhead{L$_{Br \gamma}$}  & \colhead{L$_{Pa \beta}$}    & \colhead{L$_{acc,Br \gamma}$} & \colhead{L$_{acc,Pa \beta}$}  &  \colhead{\.{M}$_{acc,Br \gamma}$}  &  \colhead{\.{M}$_{acc,Pa \beta}$} \\
   & (10$^{-4}$L$_{\sun}$) & (10$^{-3}$L$_{\sun}$) & (L$_{\sun}$) & (L$_{\sun}$) & (10$^{-8}$M$_{\sun}$y$^{-1}$) & (10$^{-8}$M$_{\sun}$y$^{-1}$) }
\startdata

070501 & 1.92 $\pm$ 0.29  & 1.83 $\pm$ 0.28  & 0.35 $\pm$ 0.05  &  1.08 $\pm$ 0.16 & 1.4 $\pm$ 0.2  & 4.2 $\pm$ 0.6 \\
070708 & 2.10 $\pm$ 0.15  & 1.19 $\pm$ 0.09  & 0.39 $\pm$ 0.03  &  0.65 $\pm$ 0.05 & 1.5 $\pm$ 0.1  & 2.5 $\pm$ 0.2  \\
080522 & 1.39 $\pm$ 0.18  & 0.87 $\pm$ 0.06  & 0.27 $\pm$ 0.03  & 0.46 $\pm$ 0.03  & 1.0 $\pm$ 0.1  & 1.8 $\pm$ 0.1 \\
090218 & 0.80 $\pm$ 0.25  & 0.77 $\pm$ 0.05  & 0.16 $\pm$ 0.05  & 0.40 $\pm$ 0.03  & 0.6 $\pm$ 0.2  & 1.5 $\pm$ 0.1 \\
090331 & 1.73 $\pm$ 0.13  & 1.12 $\pm$ 0.08  & 0.33 $\pm$ 0.02  & 0.61 $\pm$ 0.04  & 1.3 $\pm$ 0.1  & 2.4 $\pm$ 0.2  \\
090427 & 2.65 $\pm$ 0.19  & 1.11 $\pm$ 0.07  & 0.48 $\pm$ 0.03  & 0.60 $\pm$ 0.04  & 1.9 $\pm$ 0.1  & 2.4 $\pm$ 0.2  \\ 
090520 & 3.00 $\pm $ 0.22 & 1.38 $\pm$ 0.10  & 0.54 $\pm$ 0.04  & 0.77 $\pm$ 0.05  & 2.1 $\pm$ 0.2  & 3.0 $\pm$ 0.2 \\
090618 & 2.32 $\pm$ 0.16  & 1.20 $\pm$ 0.08  & 0.42 $\pm$ 0.03  & 0.66 $\pm$ 0.05  & 1.6 $\pm$ 0.1  & 2.6 $\pm$ 0.2  \\
090710 & 3.28 $\pm$ 0.23  & 1.54 $\pm$ 0.11  & 0.58 $\pm$ 0.04  & 0.88 $\pm$ 0.06  & 2.3 $\pm$ 0.2  & 3.4 $\pm$ 0.2  \\
110322 & 2.24 $\pm$ 0.13  & 1.12 $\pm$ 0.64 &  0.41 $\pm$ 0.02& 0.61 $\pm$ 0.04 &1.6 $\pm$  0.1 & 2.4 $\pm$ 0.1 \\
110323a & 3.66 $\pm$ 0.20 & 1.31$\pm$ 0.79 &  0.64  $\pm$  0.04 & 0.73$\pm$  0.04&  2.5 $\pm$  0.1 & 2.8 $\pm$ 0.2 \\
110323b & 4.19 $\pm$ 0.37 & 1.43 $\pm$1.00 &  0.72 $\pm$ 0.06 &  0.81 $\pm$ 0.06 & 2.8 $\pm$ 0.2  &  3.2 $\pm$ 0.2 \\
110428 & 3.87 $\pm$  0.27 & 1.00  $\pm$ 0.50 & 0.67 $\pm$ 0.05 &  0.54 $\pm$  0.03 & 2.6 $\pm$ 0.2 &  2.1 $\pm$ 0.1 \\

\enddata
\tablenotetext{a}{ ~The line luminosities are based on a distance of 160 pc, that of the A0V star companion SAO 206462 \citep{grady09}. The conversion of line luminosities to accretion luminosities uses the relationships derived by \citet{muzerolle98b} for Pa $\beta$ while a revised one by \citet{calvet04} was used for Br $\gamma$. These are derived for lower accretion rates and may suffer from systematic uncertainties. The quoted uncertainties do not include that of the original calibrations.}
\tablenotetext{b}{ ~Using \.{M} $\approx$ L$_{acc}$R$_{star}$/GM$_{star}$ with R$_{star}$ = 2.3 R$_{\sun}$ and M$_{star}$ = 1.6 M$_{\sun}$.}

\end{deluxetable}

\clearpage

\clearpage

\clearpage

\clearpage

\clearpage

\clearpage

\clearpage

\clearpage

\clearpage

\clearpage

\clearpage

\end{document}